\documentclass[twocolumn,amsmath,amsfonts,amssymb]{revtex4}
\usepackage{graphicx}
\def\beq{\begin{equation}}
\def\eeq{\end{equation}}
\def\bea{\begin{eqnarray}}
\def\eea{\end{eqnarray}}

\begin{document}

\title{Critical behaviors and phase transitions of black holes in higher order gravities and extended phase spaces}
\author{Zeinab Sherkatghanad}
\email{z.sherkat@ph.iut.ac.ir}
\affiliation{Department of Physics, Isfahan University of Technology, Isfahan, 84156-83111, Iran}
\author{Behrouz Mirza}
\email{B.mirza@cc.iut.ac.ir}
\affiliation{Department of Physics, Isfahan University of Technology, Isfahan, 84156-83111, Iran}
\author{Zahra Mirzaiyan}
\email{z.mirzaiyan@ph.iut.ac.ir}
\affiliation{Department of Physics, Isfahan University of Technology, Isfahan, 84156-83111, Iran}
\author{Seyed Ali Hosseini Mansoori}
\email{sa.hosseinimansoori@ph.iut.ac.ir}
\affiliation{Department of Physics, Isfahan University of Technology, Isfahan, 84156-83111, Iran}

\begin{abstract}
We consider the critical behaviors and phase transitions of Gauss Bonnet-Born Infeld-AdS black holes (GB-BI-AdS) for $d=5,6$ and the extended phase space. We assume the cosmological constant, $\Lambda$,  the coupling coefficient $\alpha$, and the BI parameter $\beta$ to be thermodynamic pressures of the system. Having made these assumptions, the critical behaviors are then studied in the two canonical and grand canonical ensembles.  We find "reentrant and triple point phase transitions" (RPT-TP) and   "multiple reentrant phase transitions" (multiple RPT) with increasing  pressure of the system  for  specific values of the coupling coefficient $\alpha$ in the canonical ensemble. Also, we observe a reentrant phase transition (RPT) of GB-BI-AdS black holes in the grand canonical ensemble and for $d=6$. These calculations are then expanded to the critical behavior of Born-Infeld-AdS (BI-AdS) black holes in the third order of Lovelock gravity and in the grand canonical ensemble to find a Van der Waals behavior for $d=7$ and a reentrant phase transition for $d=8$ for specific values of potential $\phi$ in the grand canonical ensemble. Furthermore, we obtain a  similar behavior for the limit of $\beta \to \infty$, i.e charged-AdS black holes in the third order of the Lovelock gravity. Thus, it is shown that the critical behaviors of these black holes are independent of the parameter  $\beta$ in the grand canonical ensemble.
\end{abstract}

\maketitle

\section{Introduction}	

Black holes are indeed known as the thermodynamic objects that can be  described by a physical temperature and an entropy \cite{Hawking:1971,Bekenstein:1973,Hawking:1973}.
Black hole thermodynamics continues to be one of the most important subjects in  gravitational physics.
The first attempts  to explain the instabilities of  anti-de Sitter (AdS) black holes are due to Hawking and Page in 1983 \cite{HawkingPage:1983}. They explored the existence of a specific phase transition in the phase space of the Schwarzschild AdS black hole. \\
Thermodynamics of black holes in the AdS space has attracted a lot of attention for many years due to the AdS/CFT correspondence \cite{CaldarelliEtal:2000,Hawking:1999dp,HawkingEtal:1999}.
It has been shown that the properties and critical behaviors of black holes in the AdS space are different from those of the black holes in an asymptotically flat spacetime \cite{Cardoso:2008,Wang:2006,Sekiwa:2006,Xie:2006,BelhajEtal:2012,lala:2012,Majhi:2012,HendiVahinidia:2012,Cai:2013qga,Ma:2013aqa,Chen:2013ce,Zhao:2013oza,Mo:2013sxa,Zou:2013owa,CveticGubser:1999a,TsaiEtal:2012,NiuEtal:2011,Poshteh:2013pba,Mirza:2014,Fernando:2006,cheng:2014,GubserMitra:2000a,Banerjee:2010bx,Gibbons:2004ai,ElvangEtal:2007,Emparan:2007wm,Myung:2009,olea:2008,Dias:2009iu,Hosseini:2013,PecaLemos:1999}. The critical behaviors of the black holes in the AdS space  have been studied in \cite{Dolan:2010,Dolan:2011a,Dolan:2011b,Dolan:2013,UranoEtal:2009,KastorEtal:2010,KastorEtal:2009,ChamblinEtal:1999a,ChamblinEtal:1999b} by including the cosmological constant as a thermodynamic pressure in the first law of black hole thermodynamics. In this approach, the black hole mass $M$ is replaced by enthalpy rather than by internal energy. \\
Recent studies have shown the analogy between  charged black holes in an AdS space and the Van der Waals fluid in an extended phase space \cite{ChamblinEtal:1999a,ChamblinEtal:1999b,Goldenfeld:1992}. Also, the phase diagrams of rotating black holes with single  and multiple spinnings are similar to those of  reentrant phase transition and triple point phenomena, respectively \cite{KubiznakMann:2012,GunasekaranEtal:2012,AltamiranoEtal:2013a,AltamiranoEtal:2013b,AltamiranoEtal:2013c,NarayananKumar:1994,Maslov:2004}.\\
The low energy effective action of the string theory contains both Einstein Hilbert and higher order of curvature terms. In this condition, the Gauss-Bonnet and third order Lovelock theory are the most important higher order terms in the theory of gravity. The black hole solutions and their thermodynamic quantities of the third order Lovelock theory have been investigated in \cite{Dehghani2008}.
Recently, the static black hole solutions of Gauss Bonnet-Born Infeld gravity and the third order of Lovelock-Born-Infeld gravity in the AdS space were investigated in \cite{Dehghani2008,zou:2010}.\\
There have also been efforts to explore the critical behavior and phase transitions of  black holes in higher order gravities. The critical behavior of charged AdS-Gauss-Bonnet black holes for $d=6$ demonstrated the possibility of triple point phenomena in the canonical ensemble \cite{Shao:2014}. Also, the critical behavior of higher order gravities including the Gauss-Bonnet and the third order  Lovelock gravity have been  investigated in \cite{Hao:2014,Xu:2014,MoLiu:2014,Wei:2012ui,Mann:2014,GBgrand}.\\
This paper investigates the critical behaviors and phase transitions of GB-BI-AdS black holes in the canonical ensemble for $d=5,6$. We enlarge the phase space by considering
the BI parameter and the coupling coefficient as  thermodynamic pressures.
We observe a new critical behavior dependent on the coupling coefficient $\alpha$ in the canonical ensemble in $d=6$. It is found that for  $0 \leq \alpha <13$, the system behaves similar to the standard liquid/gas of the Van der Waals fluid. For $13 \leq \alpha <16$, $18 \leq \alpha <25$ and $32 \leq \alpha <40$, the black hole admits a reentrant large/small/large black hole phase transition. For $16 \leq \alpha <18$, a reentrant  phase transition occurs for a special range of pressures while we also observe a triple point phenomenon as the pressure increases. Also, we have multiple reentrant phase transitions for  $25 \leq \alpha <32$. For $\alpha\geq 40$, there is no phase transition. We study the critical pressures with respect to the coupling coefficient $\alpha$ for these black holes.\\
The Van der Waals behavior is investigated in
 the BI-AdS black holes for $d=5,6$  in the canonical ensemble. Also, we observe a reentrant phase transition (RPT) of GB-BI-AdS black holes in the grand canonical ensemble and for $d=6$. Moreover, the critical behavior of both BI-AdS and charged-AdS black holes is investigated in the third order of Lovelock gravity in the grand canonical ensemble. We find the Van der Waals behavior for $d=7$ and a RPT for $d=8$ for the special values of potential $\phi$ in the grand canonical ensemble. Thus, the critical behaviors of these black holes are independent of the coupling coefficient  $\beta$ in the grand canonical ensemble. \\
The Einstein GB-AdS gravity has an insolated point in the space of parameters, $\alpha= l^2 / 4$. In five dimensions this point corresponds to Chern-Simons AdS gravity, which has enhanced
gauge symmetry from local Lorentz to local AdS. In higher dimensions two branches of
the solutions coincide (in pure gravity case) because the AdS vacuum becomes degenerate.
In either case, due to degeneracy, it is not allowed to take the limit  $\alpha \to l^{2}/4$. The BI field
can cancel this degeneracy, but then the Maxwell limit $\beta \to \infty$ is not well-dfined. Thus,
it is safer to restrict to $\alpha \neq l^2/4$.\\

The outline of this paper is as follows: In Sec. II,  the critical behavior and phase transitions of GB-BI-AdS black holes are examined in the canonical and grand canonical ensembles for $d=5,6$. Also, we study the BI-AdS black holes for $d=5,6$ and in the canonical ensemble.
In Sec. III, the critical behavior and phase transitions of  BI-AdS and charged-AdS black holes in the third order of the Lovelock gravity are investigated for $d=7,8$  in the grand canonical ensemble.

\section{Gauss-Bonnet-Born-Infeld-AdS  Black Holes}

\label{sec:1}
The action of the Einstein Gauss-Bonnet gravity in the presence of a nonlinear Born-Infeld electromagnetic field with a negative cosmological constant reads as follows \cite{zou:2010}:
\begin{eqnarray}
 \label{actionGB}
 I=\frac{1}{16\pi}\int d^{n+1} x\sqrt{-g} \ [R-2\Lambda +\tilde{\alpha} \ {\cal{L}}_{GB}+\cal{L}_F\ ],
\end{eqnarray}
where, $\tilde{\alpha}$  is the Gauss-Bonnet coefficient and $\Lambda=-\frac{n (n-1)}{2 l^2}$ is a negative cosmological constant. \\
The Gauss-Bonnet Lagrangian and $\cal{L}_F$ are given by
\begin{eqnarray}
 \label{EGB}
&&{\cal{L}}_{GB}=R_{\mu \nu \delta \gamma} R^{\mu \nu \delta \gamma }-4 R_{\mu \nu} R^{\mu \nu}+R^2,\\
&&{\cal{L}_{F}} =4 \beta ^2 (1-\sqrt{1+\frac{ F^{\mu \nu} F_{\mu \nu }}{2 \beta^2}}) \ ,
\end{eqnarray}
where, the constant $\beta$ is the BI parameter and the Maxwell field strength is defined by  $F_{\mu \nu}=\partial _{\mu} A _{\nu}-\partial _{\nu } A_{\mu}$
with $A_{\mu}$ as the vector potential. Let us consider the following metric:
 \begin{eqnarray}
 \label{metric}
  ds^2&=&-f(r) d t^2+\frac{1}{f(r)} d r^2+r^2 h_{ij} dx ^i dx^j,
\end{eqnarray}
where, $h_{ij}$ is a ($n-1$)-dimensional hypersurface. The metric coefficient $f(r)$ for static  GB-BI-AdS black holes is given by
 \begin{eqnarray}
 \label{MF}
f(r)=1 \mp \frac{r^2}{2\alpha }(1-\sqrt{g(r)}),
\end{eqnarray}
where, $g(r)$ is
\begin{eqnarray}
 \label{gGB}
g(r)&=&1-\frac{4 \alpha}{l^2}+\frac{4 \alpha m}{r^n}-\frac{16 \alpha \beta^2}{n(n-1)}-\frac{8 (n-1)\alpha q^2}{n \ r^{2n-2}} \\\nonumber
&\times& _2 F_1 [\frac{n-2}{2n-2},\frac{1}{2},\frac{3n-4}{2n-2},-\frac{(n-1) (n-2) q^2}{2 \beta ^2 r^{2 n-2}}]\\\nonumber
&+&\frac{8\sqrt{2} \alpha \beta \alpha}{n(n-1) r^{n-1}}\sqrt{2 \beta^2 r^{2 n-2} +(n-1)(n-2) q^2},\\\nonumber
\end{eqnarray}
here,
\begin{eqnarray}
&&\alpha=(n-2)(n-3) \tilde{\alpha},\\\nonumber
&&Q=\frac{q \sum _{n-1}}{4 \pi} \sqrt{\frac{(n-1)(n-2)}{2}},
\end{eqnarray}
and $m$ is an integration constant which is related to mass
\begin{eqnarray}
\label{massGB}
M&=&\frac{(n-1) \sum _{n-1} m}{16 \pi}\\\nonumber
&=&\frac{(n-1) \sum_{n-1} }{16 \pi  } \Bigl[\alpha k^2  r_+^{n-4}+k r_+^{n-2}\\\nonumber
&+&\frac{64 \pi^2 Q^2 r_+^{2-n}}{ n (n-2)} \, _2F_1[\frac{1}{2},\frac{n-2}{2 n-2};\frac{3 n-4}{2 n-2};-\frac{16 \pi ^2 Q^2 r_+^{2-2 n}}{\beta ^2}]\\\nonumber
&&+\frac{2 r_+^n}{n \ (n-1)}(-2 \beta ^2 \sqrt{\frac{16 \pi^2 Q^2 r_+^{2-2 n}}{\beta ^2}+1}+2 \beta ^2+8 \pi  P)\Bigr],
\end{eqnarray}
where, $\sum\nolimits_{n-1}$ exhibits the volume of the constant curvature hypersurface described by $h_{ij} dx ^i dx^j$ and $r_+$ is the horizon radius of the black hole determined by the largest real root of $f(r_+)=0$. We study only the Einstein branch of the black holes, i.e., with the negative sign in Eq. (\ref{MF}). This branch has well-difined limit $\alpha \ \to 0$ that leads to Einstein-Hilbert-BI black holes. The second branch \cite{BoulwareDeser} has not been considered. 
In the following calculations, we consider $\sum\nolimits_{n-1}=1$ and the specific case $k=1$ for simplicity. The thermodynamic quantities of these black holes are:
\begin{eqnarray}
\label{thermodynamic}
T&=&\frac{1}{12 \pi  r_+ \left(2 \alpha +r_+^2\right)}\Bigl(3 \alpha  (n-4)+\frac{48 \pi  P r_+^4}{(n-1)} \\\nonumber
&+&\frac{12 \beta ^2 r_+^4}{(n-1)} \left(1-\sqrt{\frac{16 \pi^2 Q^2 r_+^{2-2 n}}{\beta ^2}+1}\right)+3  (n-2) r_+^2 \Bigr),\\
\label{entropy}
S&=&\frac{(n-1) r_+^{n-5} }{4} \left(\frac{2 \alpha   r_+^2}{n-3}+\frac{r_+^4}{n-1}\right),\\
\label{potential}
\phi&=&\frac{4 \pi Q}{(n-2) r_+ ^{n-2}} \\\nonumber
&\times& \, _2F_1[\frac{1}{2},\frac{n-2}{2 n-2},\frac{3 n-4}{2 n-2},-\frac{16 \pi ^2 Q^2 r_+^{2-2 n}}{\beta ^2}],
\end{eqnarray}
where, $n=d-1$ and we set $P=-\frac{1}{8\pi }\Lambda =\frac{(d-1)(d-2)}{16\pi l^{2}}$.
The first law of thermodynamics and The Smarr relation for GB-BI black holes take the following form:
 \begin{eqnarray}
 \label{7b}
dH&=&TdS+\phi dQ+V dP+{\cal{A}}  d\alpha+{\cal{B}}  d \beta,\\
H&=&\frac{d-2}{d-3} TS + Q \phi- \frac{2}{d-3} P V -\frac{2}{d-3} \alpha {\cal{A}}\\\nonumber
&-&\frac{1}{d-3}\beta {\cal{B}},
\end{eqnarray}
where, $H=M$ is the enthalpy of the gravitational system \cite{ChamblinEtal:1999a,ChamblinEtal:1999b,Dolan:2012}.\\
The parameters $V$, $\cal{A}$, and $\cal{B}$ are the thermodynamic quantities conjugating to  pressure $P$, Gauss Bonnet coupling coefficient $ \alpha$, and Born-Infeld parameter $\beta$, respectively. These parameters are determined from either  the first law of thermodynamics or the Smarr relation:
\begin{eqnarray}
 \label{7b}
V&=&\frac{r_+^{d-1}}{d-1}=\frac{1}{d-1}(\frac{(d-2) \it{v}}{4})^{d-1},\\
{\cal{A}}&=&\frac{(d-2) r_+^{d-6}}{16 \pi }-\frac{d-2}{2 (d-4)} T r_+^{d-4},\\
{\cal{B}}&=&\frac{r_+^{d-1}}{2  \pi  \beta  (d-1)} \Bigl(\beta ^2 -\beta ^2  \sqrt{\frac{16 \pi^2 Q^2 r_+^{4-2 d}}{\beta ^2}+1}\\\nonumber
&+&\frac{8 \pi^2 Q^2 r_+^2}{r_+^{2 d-2} } \, _2F_1[\frac{1}{2},\frac{d-3}{2 d-4},\frac{7-3 d}{4-2 d},-\frac{16 \pi^2 Q^2 r_+^{4-2 d}}{\beta ^2}]
\Bigr),
\end{eqnarray}
Here, $\it{v}$ is the effective specific volume.\\

In the following section, we will investigate the critical behaviors and phase transitions of the  GB-BI-AdS black hole in the canonical ensembles.
%%%%%%%%%%%%%%%%%%%%%%%%%%%%%%%%%%%%%%%%%%%%%%%

\subsection{Critical behavior in the canonical ensemble }

The phase transitions and critical exponents of the BI-AdS black holes for $d=4$ were calculated in the canonical and grand canonical ensembles \cite{GunasekaranEtal:2012,Fernando:2006}. Also, the stability analysis of five dimensional GB-BI black holes in the AdS space were studied in \cite{zou:2010}. \\
Now,  let us investigate the critical behavior of GB-BI-AdS black holes in the canonical ensemble and the extended phase space for $d=5,6$. In the canonical ensemble, we adopt fixed values for $\beta$, $Q$,  and $\alpha$ and consider the $P-v$ extended phase space.\\
Using Eq. (\ref{thermodynamic}), we obtain the following equation of state for the black hole system in the canonical ensemble:
\begin{eqnarray}
\label{equationstate}
P&=&\frac{T}{v} -\frac{(d-3)}{(d-2) \pi v^2}+\frac{32 T \alpha}{(d-2)^2 v^3}\\\nonumber
&&-\frac{16 \alpha  (d-5)}{(d-2)^3 \pi v^4} +\frac{ \beta ^2}{4 \pi} \sqrt{\frac{16 \pi^2 Q^2 (d-2)^{4-2 d} v^{4-2 d}}{\beta ^2}+1}\\\nonumber
&&-\frac{ \beta ^2}{4 \pi}.
\end{eqnarray}
Thus, the critical points can be determined by using the following conditions
\begin{eqnarray}\label{critical}
\frac{\partial P}{\partial v}=0, \ \ \frac{\partial ^2 P}{\partial v ^2}=0 \ .
\end{eqnarray}
We are taking into acount the above conditions to determine the critical temperature $T_c$ and volume $V_c$ for the specific values of $Q$, $\beta$ and $\alpha$, then we replace these critical parameters in the equation of state, Eq. (\ref{equationstate}) to find the critical pressures.
Also, in the canonical ensemble, the Gibbs free energy  from Eqs. (\ref{massGB}), (\ref{thermodynamic}), and (\ref{entropy}) for fixed values of $\beta$, $Q$, and $\alpha$ for $d=6$ is given by
\begin{eqnarray}
\label{gibbs}
G&=&M-T S\\\nonumber
&=&\frac{3 r_+^4}{240 \pi  r_+^3 \left(2 \alpha +r_+^2\right)}\Bigr(20 \alpha ^2-5 \alpha  r_+^2-4 \pi  P r_+^6\\\nonumber
&-&48 \pi \ \alpha  P r_+^4+ (\beta ^2 r_+^6+12 \alpha  \beta ^2 r_+^4 )  \sqrt{\frac{16 \pi^2 Q^2}{\beta ^2 r_+^8}+1} -\beta ^2 r_+^6\\\nonumber
&-&12 \alpha  \beta ^2 r_+^4+5 r_+^4\Bigl)+\frac{256 \pi^2 Q^2 }{240 \ \pi  r_+^3} \, _2F_1[\frac{3}{8},\frac{1}{2},\frac{11}{8},-\frac{16 \pi^2 Q^2}{r_+^8 \beta ^2}].
\end{eqnarray}
The critical behavior of the Gibbs free energy with respect to  temperature depends on the values of the coupling coefficients $\alpha$ in the canonical ensemble.\\
In what follows, we discuss in detail some interesting features of the critical behavior of GB-BI-AdS black holes depending on the coupling coefficient $\alpha$ for $\beta=1$ and $Q=1$.\\

\subsubsection{ Van der Waals behaviour}

\begin{figure}
\begin{center}
  \includegraphics[width=7cm]{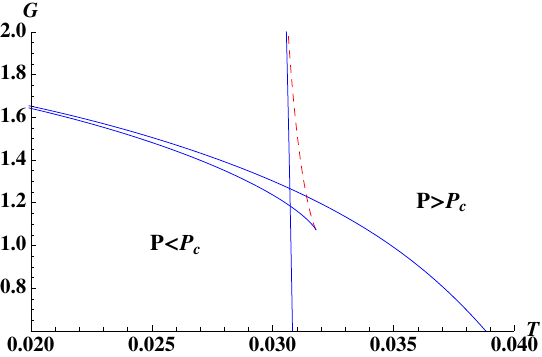}
  \caption[] {\it{Gibbs free energy $G$ with respect to the temperature
   $T$ of GB-BI-AdS black holes for $d=6$, $Q=1$, $\beta=1$ and $\alpha=12$. The  dashed (red) and solid (blue) lines
correspond to the negative and positive $C_{P}s$, respectively. At $P<P_c$, the black hole experiences a first order phase transition. There is no phase transition at $P>P_c$.}}\label{figure:GT}
\end{center}
 \end{figure}

For $0 \leq \alpha <13$, the critical behavior of the Gibbs free energy with respect to  temperature is depicted in Fig. \ref{figure:GT} for fixed values of $\beta$, $Q$, $\alpha$ and $d=6$.\\
\begin{figure}
\begin{center}
  \includegraphics[width=7cm]{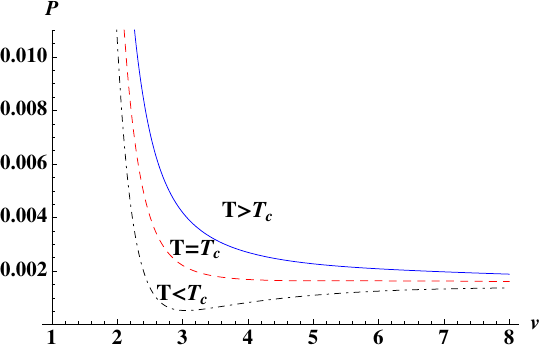}\\
  \caption {\it{The P-v diagram of GB-BI-AdS black hole  for $d=6$, $Q=1$, $\beta=1$, and $\alpha=12$. The temperature of isotherms increase from bottom to top.
The upper (blue) solid line correspond to the “ideal gas”, we have one phase for $T > Tc$, the  critical isotherm $T = Tc$ is
denoted by the dashed (red)  line. The lower dot-dashed (black)  line corresponds to
temperatures smaller than the critical temperature $T <Tc$.}}\label{figure:PV}
\end{center}
 \end{figure}
In this case, we observe one critical point at $P =P_c$ and the swallowtail behavior, i.e, a first order phase transition between small and large black holes, for $P < P_c$.  Thus, a "standard liquid/gas"  Van der Waals phase transition occurs in this limit. The corresponding $P-v$ diagram is displayed in Fig. \ref{figure:PV}. \\

Also, we consider the limit of $\alpha=0$, i.e the Born-Infeld-AdS black holes.  In this condition, a Van der Waals behavior for $d=5,6$ and $Q=1$ and all the values of $\beta$ is investigated which is similar to that in Fig. \ref{figure:GT}. Furthermore, repeating our calculations for GB-BI-AdS for $d=5$, we will see  a Van der Waals behavior  for $Q=1$ and all the ranges of $\beta$ and $\alpha$.

\begin{figure*}
\centering
\begin{tabular}{ccc}
\rotatebox{0}{
\includegraphics[width=0.29\textwidth,height=0.24\textheight]{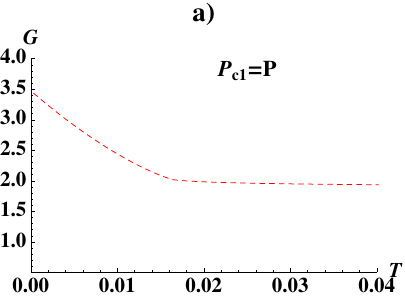}}&
\rotatebox{0}{
\includegraphics[width=0.29\textwidth,height=0.24\textheight]{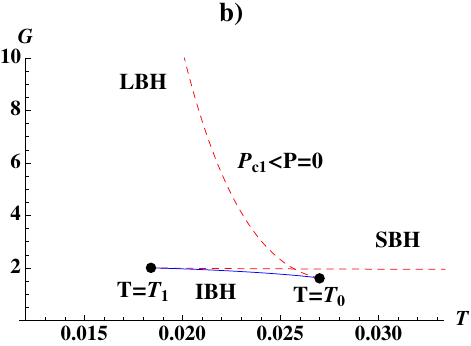}}&
\rotatebox{0}{
\includegraphics[width=0.29\textwidth,height=0.24\textheight]{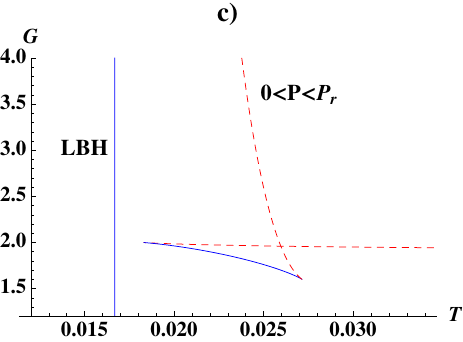}}\\
\rotatebox{0}{
\includegraphics[width=0.29\textwidth,height=0.24\textheight]{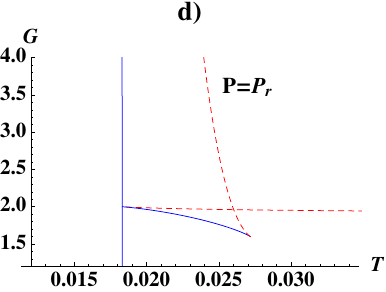}}&
\rotatebox{0}{
\includegraphics[width=0.29\textwidth,height=0.24\textheight]{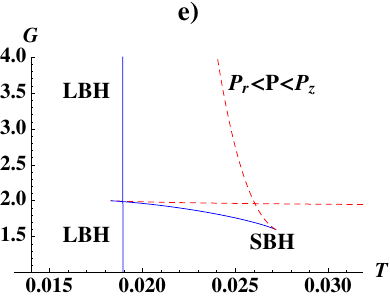}}&
\rotatebox{0}{
\includegraphics[width=0.31\textwidth,height=0.24\textheight]{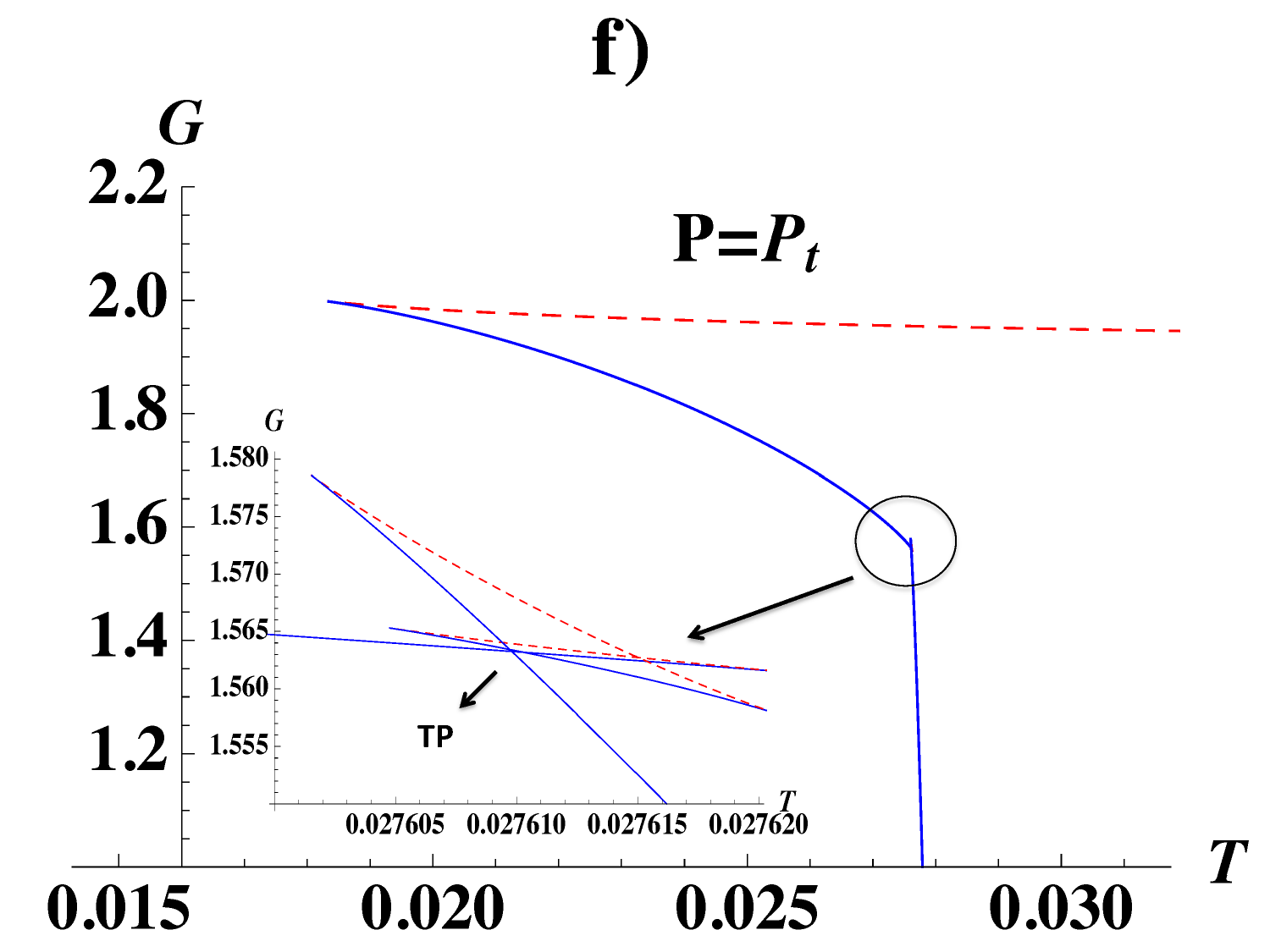}}\\
\rotatebox{0}{
\includegraphics[width=0.3\textwidth,height=0.26\textheight]{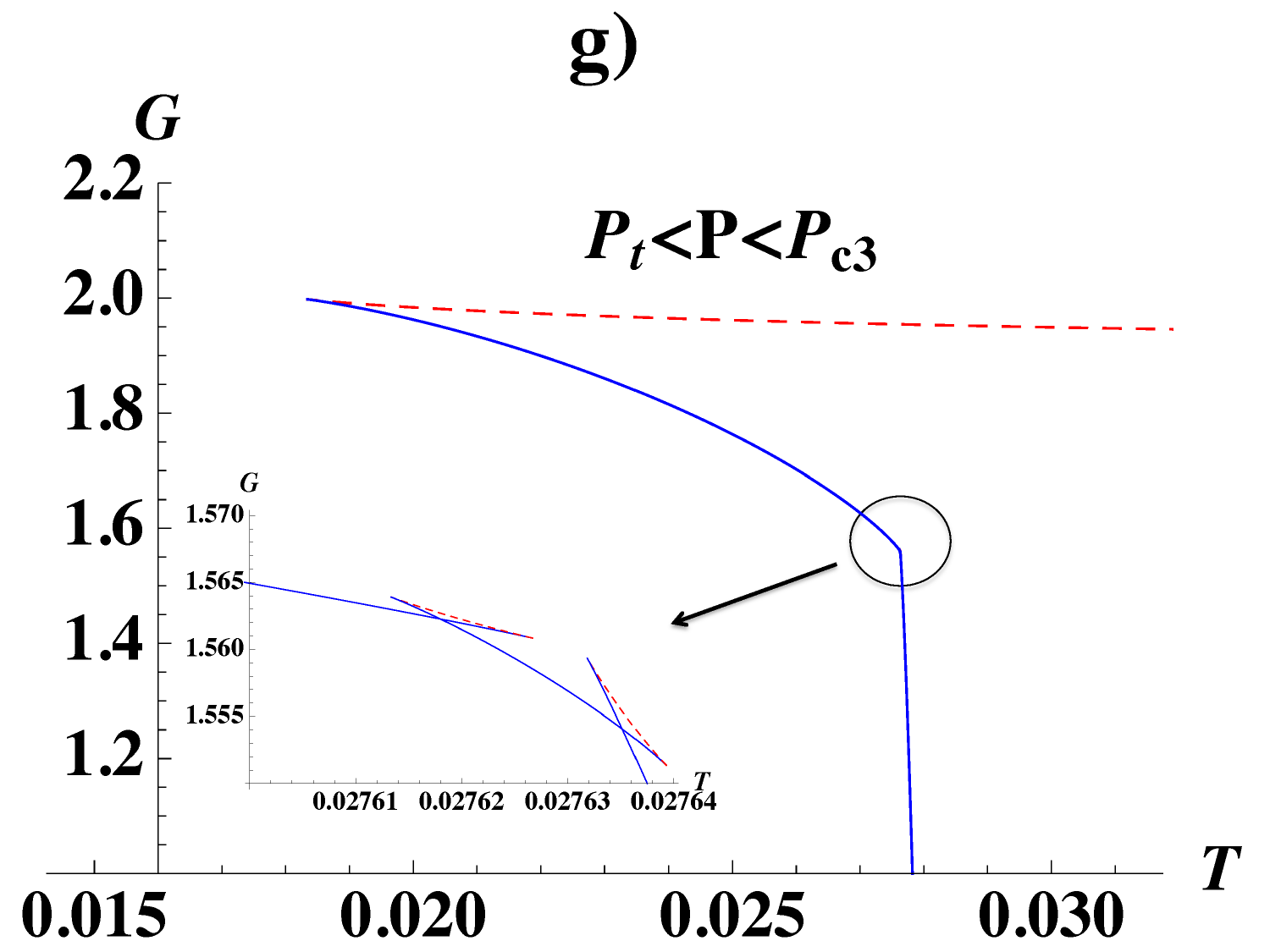}}&
\rotatebox{0}{
\includegraphics[width=0.3\textwidth,height=0.26\textheight]{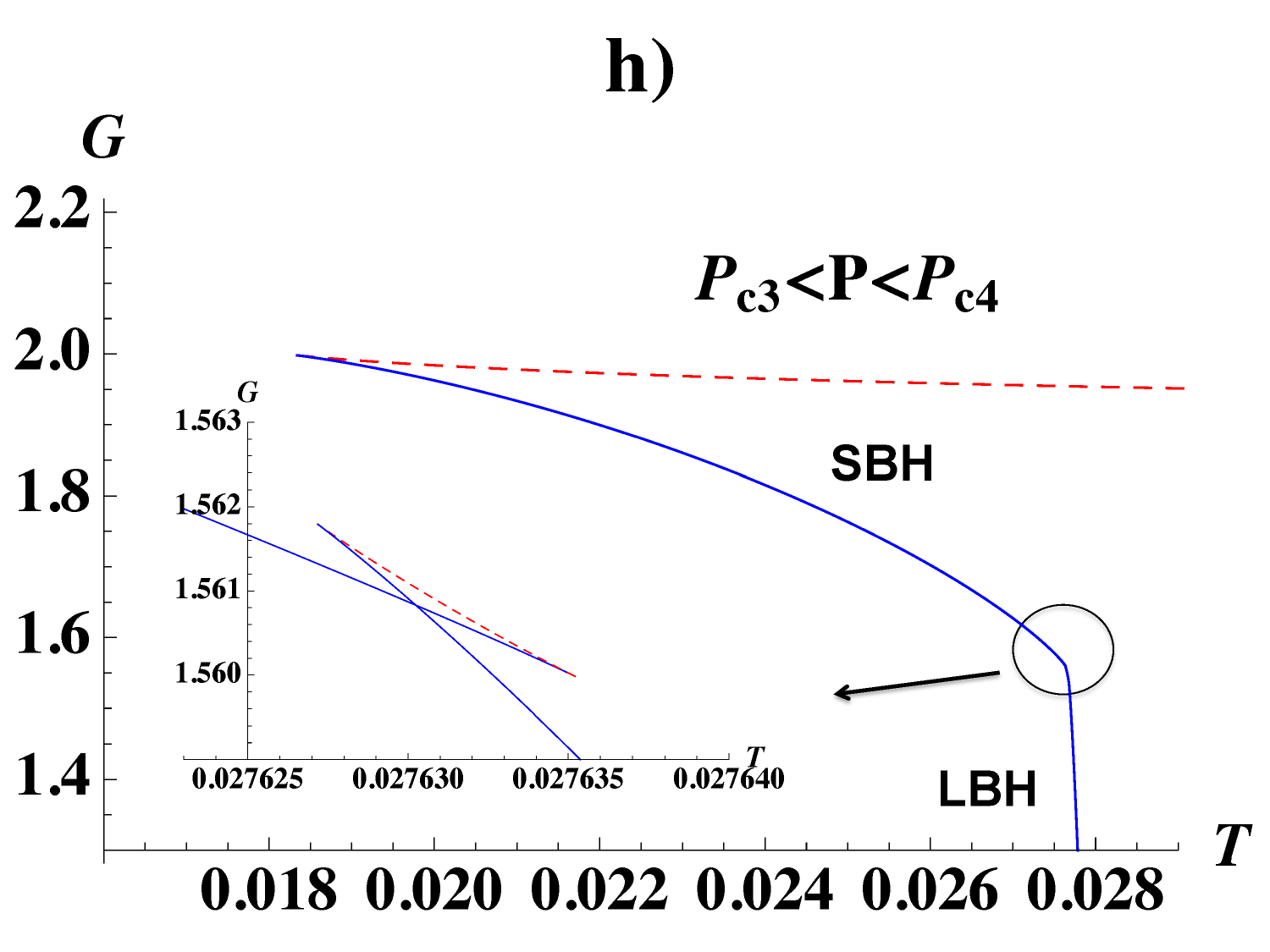}}&
\rotatebox{0}{
\includegraphics[width=0.3\textwidth,height=0.26\textheight]{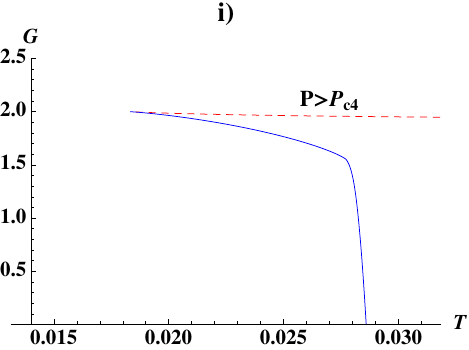}}\\
\\
%width=0.4\textwidth,height=0.33\textheight
\end{tabular}
\caption{ Reentrant-Triple point phase transition for $d=6$, $Q=1$, $\beta=1$ and $\alpha=\frac{33}{2}$ of GB-BI-AdS black holes. Here, the Gibbs free energy is displayed with respect to temperature for  various values of pressures $P=\{-0.061, 0, 0.0003, 0.00037, 0.0004, 0.001181, 0.001188,0.00122, 0.0013\}$ (from top to bottom). We consider Solid (blue)/dashed (red) lines corresponding to positive and negative $C_P$, respectively.
Picture a) shows the thermodynamically unstable branch of $P_{c1}=-0.061$ with a negative $C_P$. In picture b), for $-0.061<P\leq 0$, we have two zero order phase transitions at $T_1=0.0183$ and $T_0=0.027$, which minimize the Gibbs free energy. In this case, the minimum value of Gibss free energy is discontinuous and we have three separate phases of large, intermediate, and small size black holes. These  phases LBH/IBH/SBH are connected by a jump in the Gibbs free energy $G$, or two ‘zero-order phase transitions’ at $T_1$ and $T_0$, respectively. In Picture c), for $0<P \leq P_r=0.00037$, we find a new branch of large thermodynamically stable black holes, at which two zero-order phase transitions at $T_1$ and $T_0$ do not minimize the Gibbs free energy. Thus, only one phase of the large black holes exists. In Picture d), at $P=P_r=0.00037$, the swallowtail in the unstable branch touches the lower stable branch, and the reentrant phase transition appears. In Picture e), for $P_{r} < P <P_{z}=0.000415$, the minimum of Gibss free energy is discontinuous. In this range of pressure, we have two phases of intermediate and small sizes black holes. These phases are connected by a jump in $G$, or a ‘zero-order phase transition’. This critical behavior admits a reentrant large/small/large black hole phase transition. At $P=P_z=0.000415$, the RPT disappears but the first order phase transition is still present. Increasing  pressure, in Picture f), at the triple point (TP), $P=P_t=0.001181$, we observe two swallowtails in the stable phase of black holes. In Picture g, h), one of these swallowtails starts from $P=P_{c1}=-0.061$ and terminates at $P= P_{c3}=0.001197$ and the other  occurs at $P=P_{c2}=0.00115$ and disappears at  $P=P_{c4}=0.00123$ although the critical point $P=P_{c2}$, does not minimize G and it is an unphysical critical point. In Picture i), above the critical point $P=P_{c4}$ the system displays no phase transitions.
}\label{figure:RETR}
\end{figure*}
\begin{figure*}
\centering
\begin{tabular}{cc}
\rotatebox{0}{
\includegraphics[width=0.4\textwidth,height=0.27\textheight]{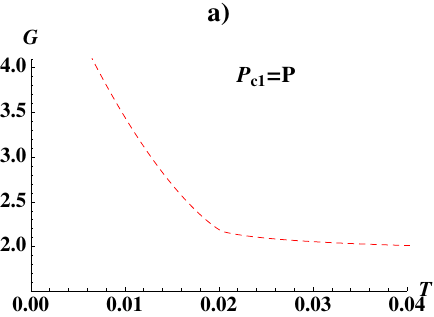}}&
\rotatebox{0}{
\includegraphics[width=0.4\textwidth,height=0.27\textheight]{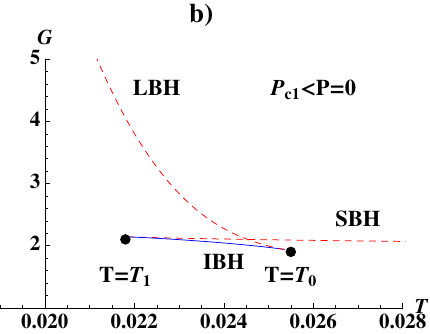}}\\
\rotatebox{0}{
\includegraphics[width=0.4\textwidth,height=0.26\textheight]{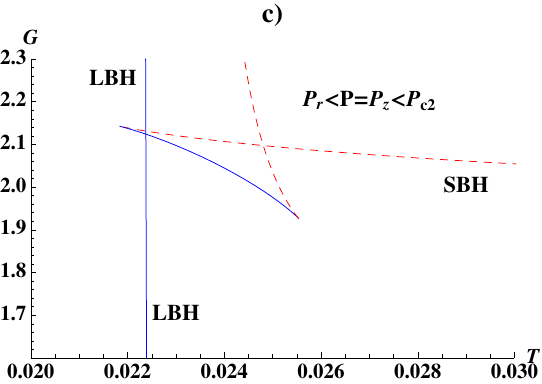}}&
\rotatebox{0}{
\includegraphics[width=0.4\textwidth,height=0.30\textheight]{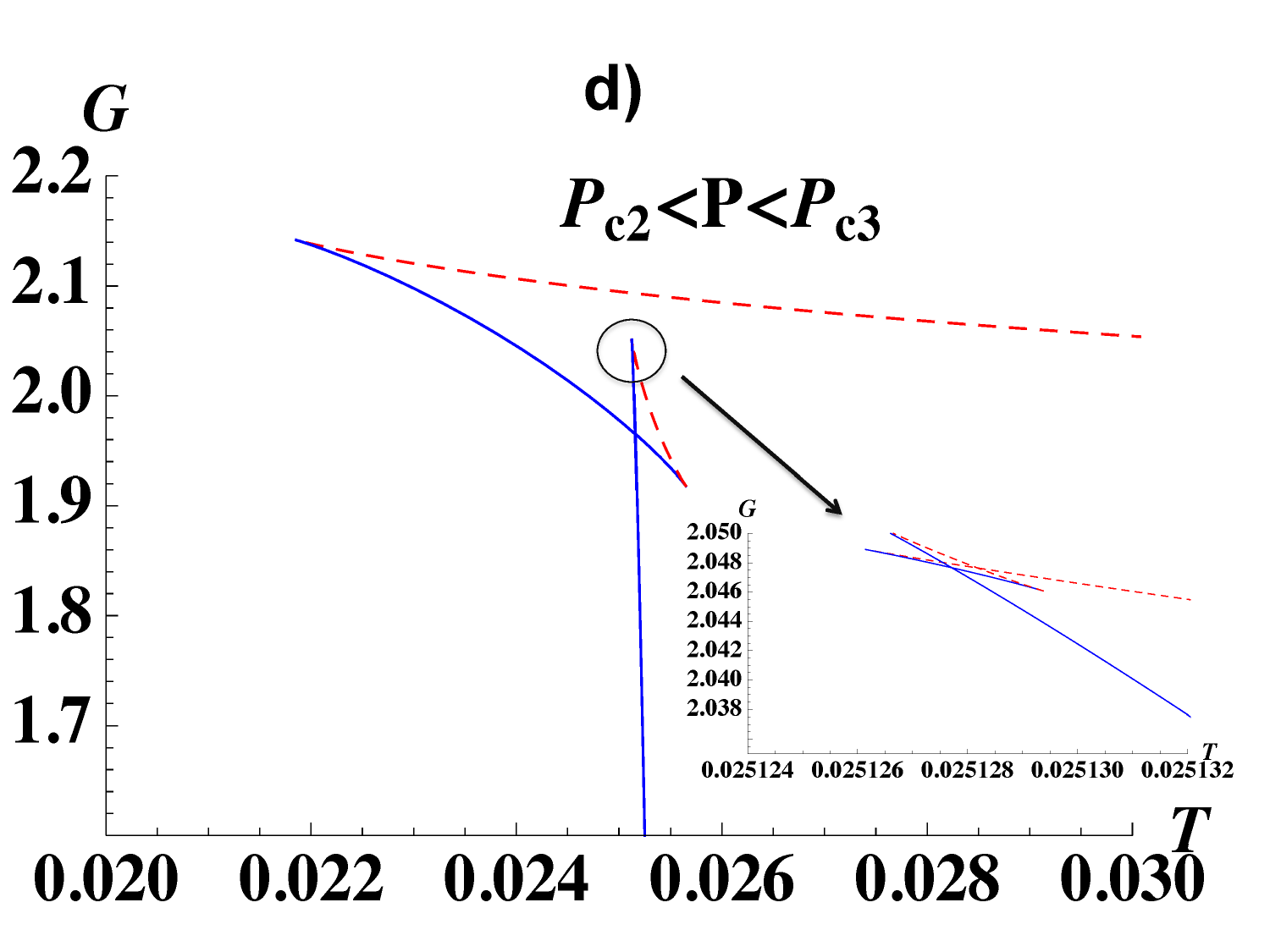}}\\
\rotatebox{0}{
\includegraphics[width=0.4\textwidth,height=0.26\textheight]{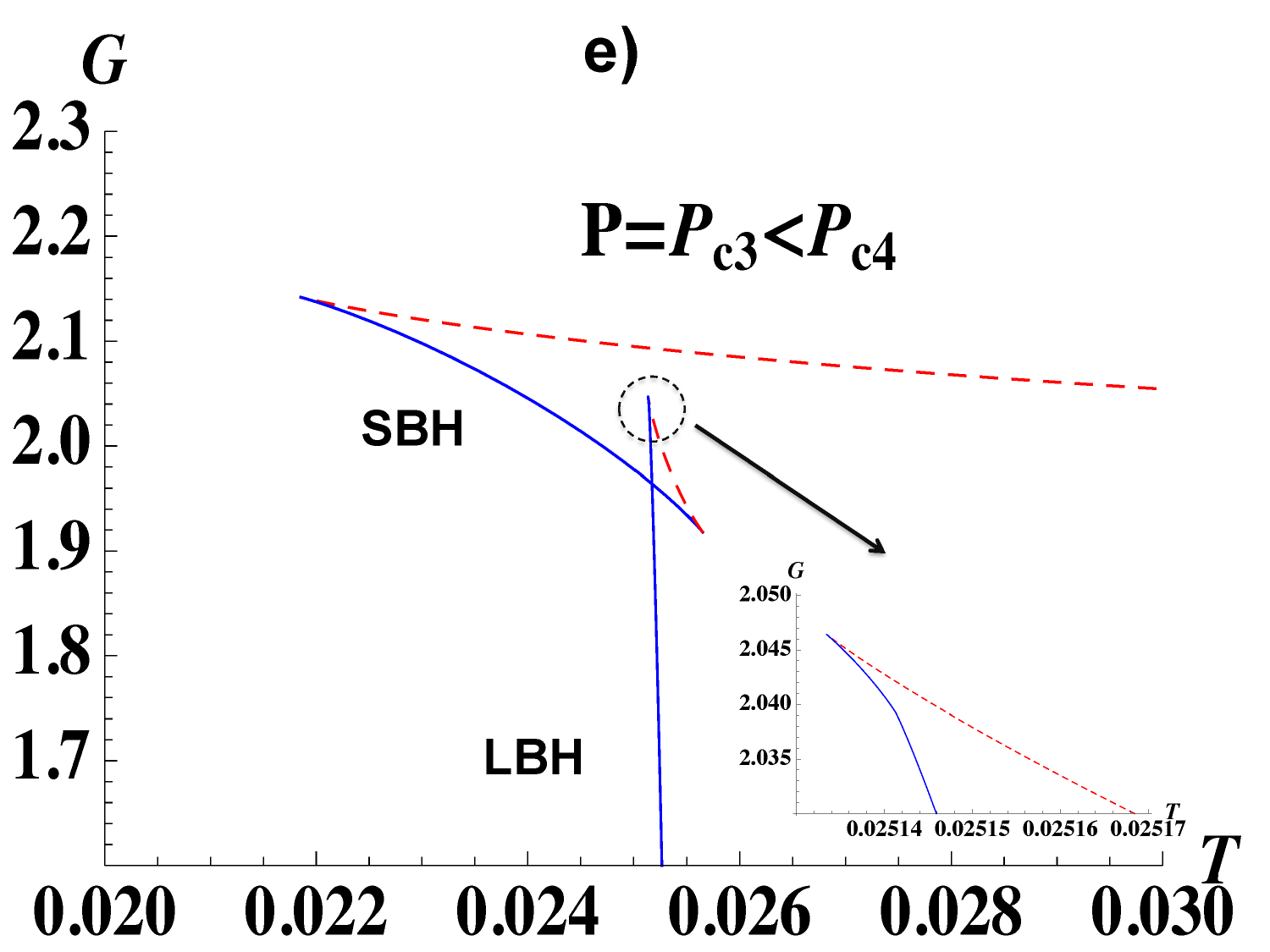}}&
\rotatebox{0}{
\includegraphics[width=0.4\textwidth,height=0.26\textheight]{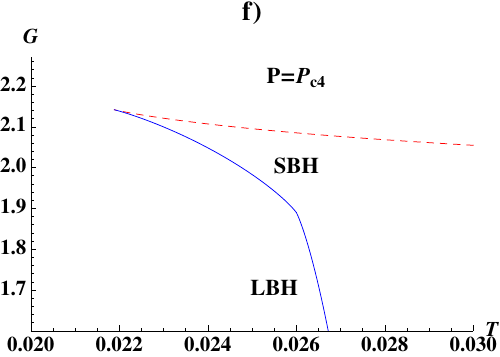}}\\
\\
%width=0.4\textwidth,height=0.33\textheight
\end{tabular}
\caption{ Reentrant phase transition for $d=6$, $Q=1$, $\beta=1$ and $\alpha=20$ of the GB-BI-AdS black holes. Here, the Gibbs free energy is displayed with respect to temperature for various values of pressures $P=\{-0.032, 0,0.00064,0.000986,0.0009897,0.00197\}$ (from top to bottom). We observe that the Solid (blue)/dashed (red) lines correspond to  positive/negative $C_P$ respectively.
At $P_{c1}=-0.032<0$, we have the thermodynamically unstable branch  with a negative $C_P$. When  pressure is increased, for $P=0$, two zero order phase transitions apear which minimize the Gibbs free energy at $T_1=0.0218$ and $T_0=0.0253$. In this situation, three phases LBH/IBH/SBH are connected by two zero-order phase transitions at $T_1$ and $T_0$, respectively. Similar to Fig. \ref{figure:RETR}, at $P_r=0.0006$, the swallowtail in the unstable branch touches the lower stable branch, and we observe the reentrant large/small/intermediate black hole phase transition for $P_{r} < P <P_{z}$. At $P_z=0.00064$, the reentrant phase transition disappearing but the first order phase transition being still present. Increasing  pressure, we observe two swallowtails. One which does not minimize the Gibbs free energy starts from $P_{c2}=0.000981$ and terminates at $ P_{c3}=0.0009897$. Another   that occurs at $P=P_{c1}$ and disappears at $P_{c4}=0.00197$ minimize the Gibbs free energy.  We, therefore,  have only two critical points. Above the critical point $P=P_{c4}$, the system displays no phase transitions.
}\label{figure:RE}
\end{figure*}
\begin{figure*}
\centering
\begin{tabular}{cccc}
\rotatebox{0}{
\includegraphics[width=0.23\textwidth,height=0.19\textheight]{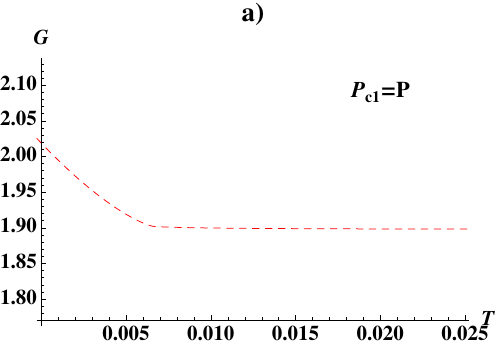}}&
\rotatebox{0}{
\includegraphics[width=0.23\textwidth,height=0.19\textheight]{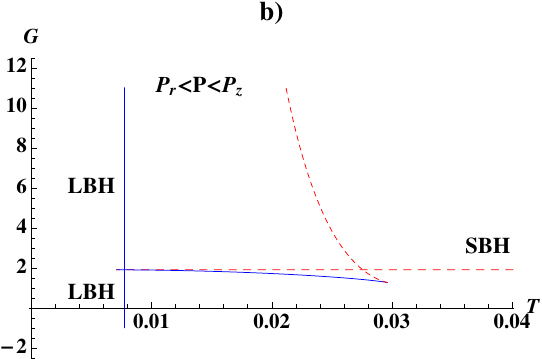}}&
\rotatebox{0}{
\includegraphics[width=0.23\textwidth,height=0.19\textheight]{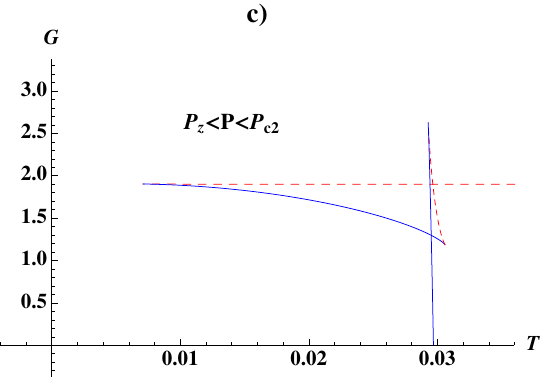}}&
\rotatebox{0}{
\includegraphics[width=0.23\textwidth,height=0.19\textheight]{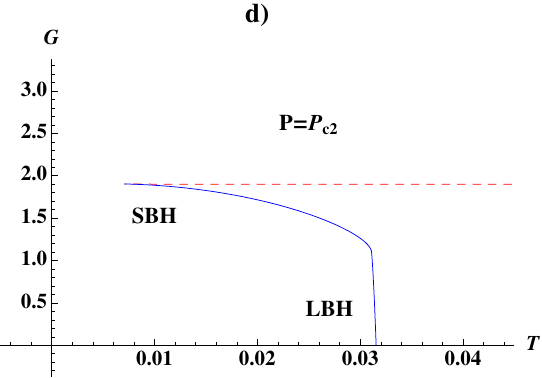}}\\
\\
%width=0.4\textwidth,height=0.33\textheight
\end{tabular}
\caption{Reentrant phase transition for $d=6$, $Q=1$, $\beta=1$ and $\alpha=13$ of GB-BI-AdS black holes. Here the Gibbs free energy is displayed with respect to temperature for various values of pressure $P=\{-0.2, 0.00006,0.0012,0.0015\}$ (from left to right). The Solid (blue)/dashed (red) lines correspond to $C_P$ positive and negative, respectively. At $P_{c1}=-0.2$, we have the thermodynamically unstable branch. For $0<P <P_{r}=0.00005$, we find a new branch of large thermodynamically stable black holes. We observe the reentrant large/small/intermediate black hole phase transition for $P_{r} \leq P <P_{z}=0.00008$. For $P_{z}<P<P_{c2}=0.0015$, we have a first order phase transition which disappears at the critical point $P_{c2}=0.0015$. Above the critical point $P=P_{c2}$, the system displays no phase transitions.
}\label{figure:RE0}
\end{figure*}
\begin{figure*}
\centering
\begin{tabular}{ccc}
\rotatebox{0}{
\includegraphics[width=0.45\textwidth,height=0.27\textheight]{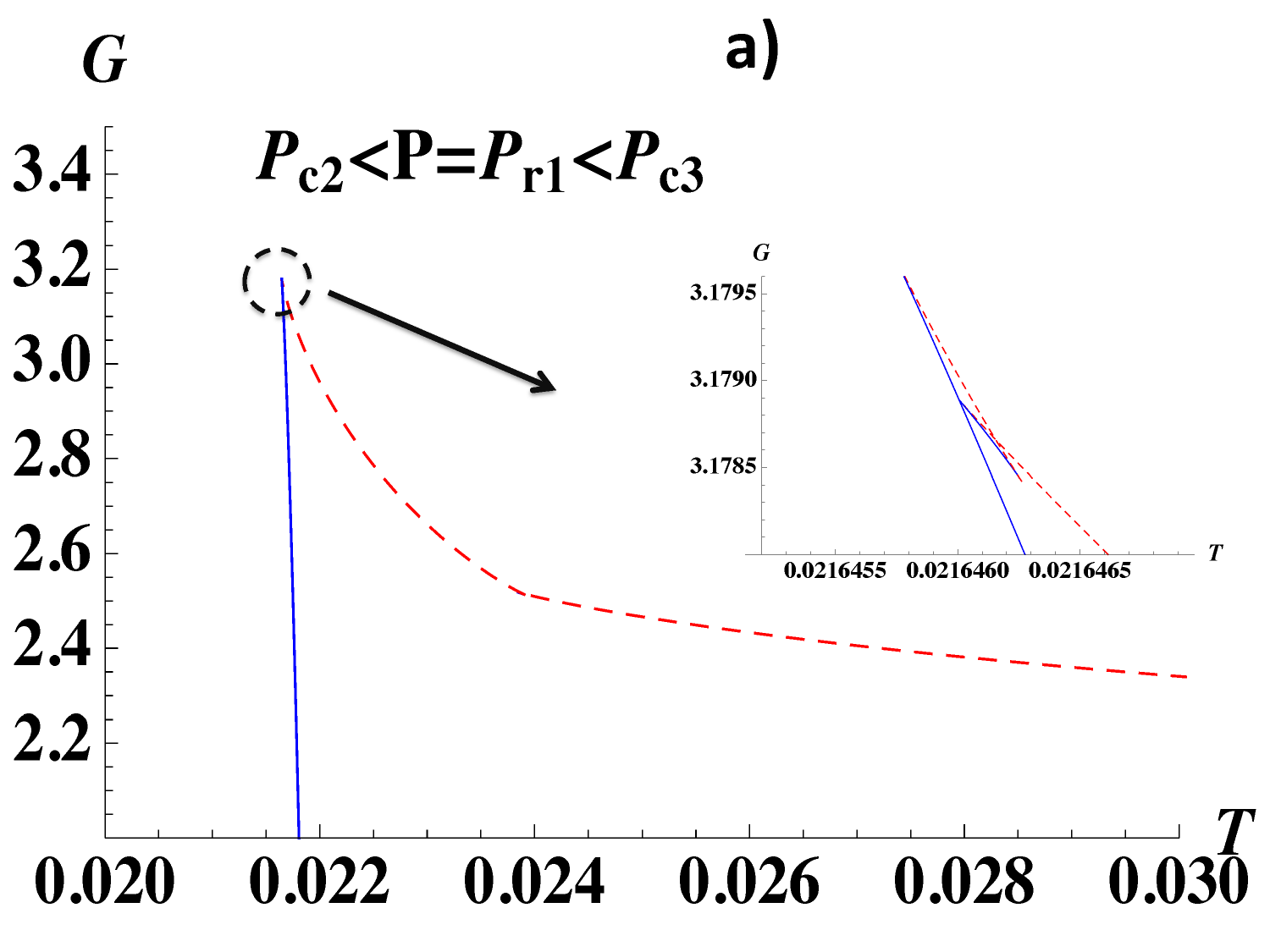}}&
\rotatebox{0}{
\includegraphics[width=0.45\textwidth,height=0.27\textheight]{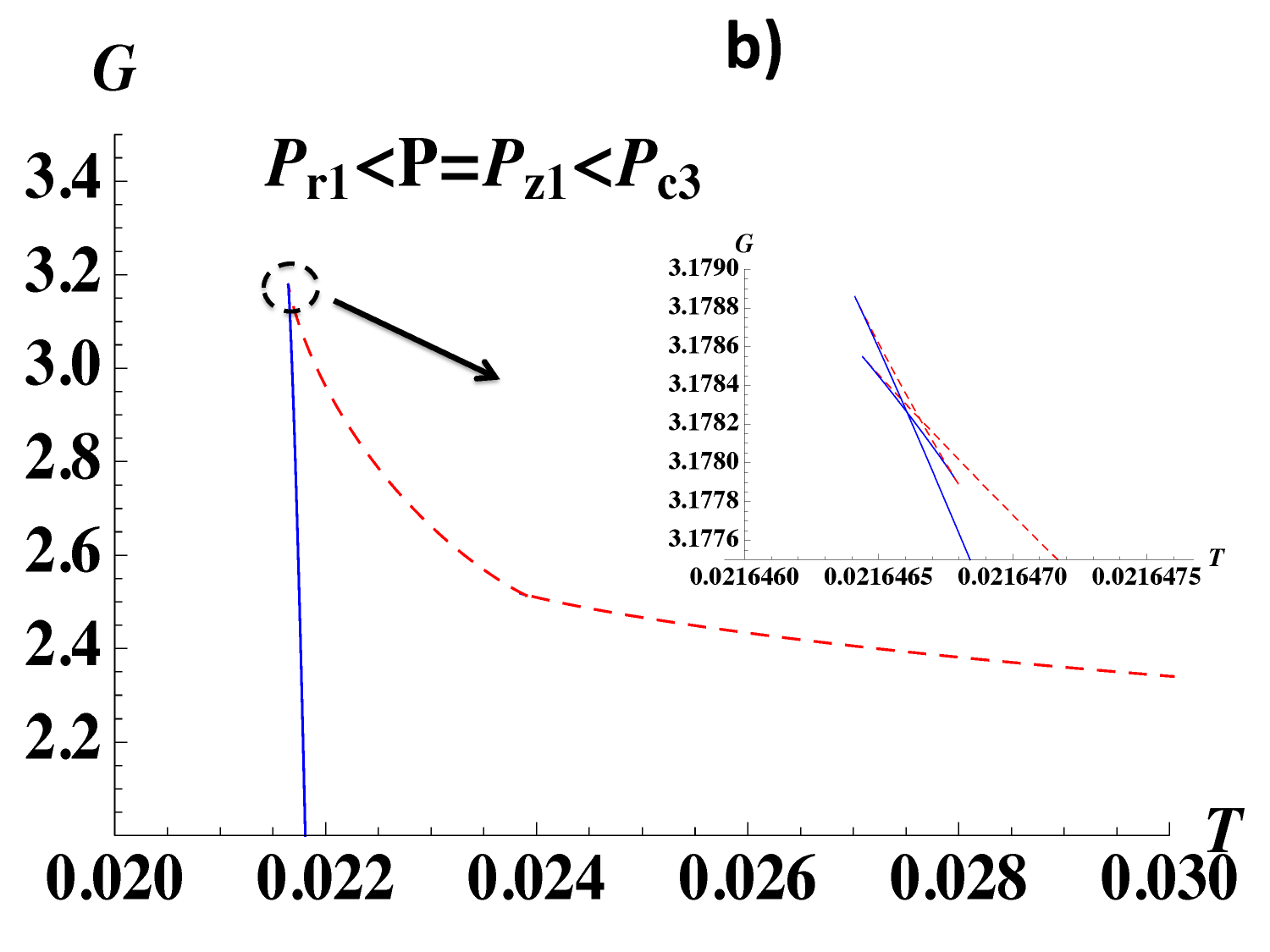}}\\
\rotatebox{0}{
\includegraphics[width=0.45\textwidth,height=0.27\textheight]{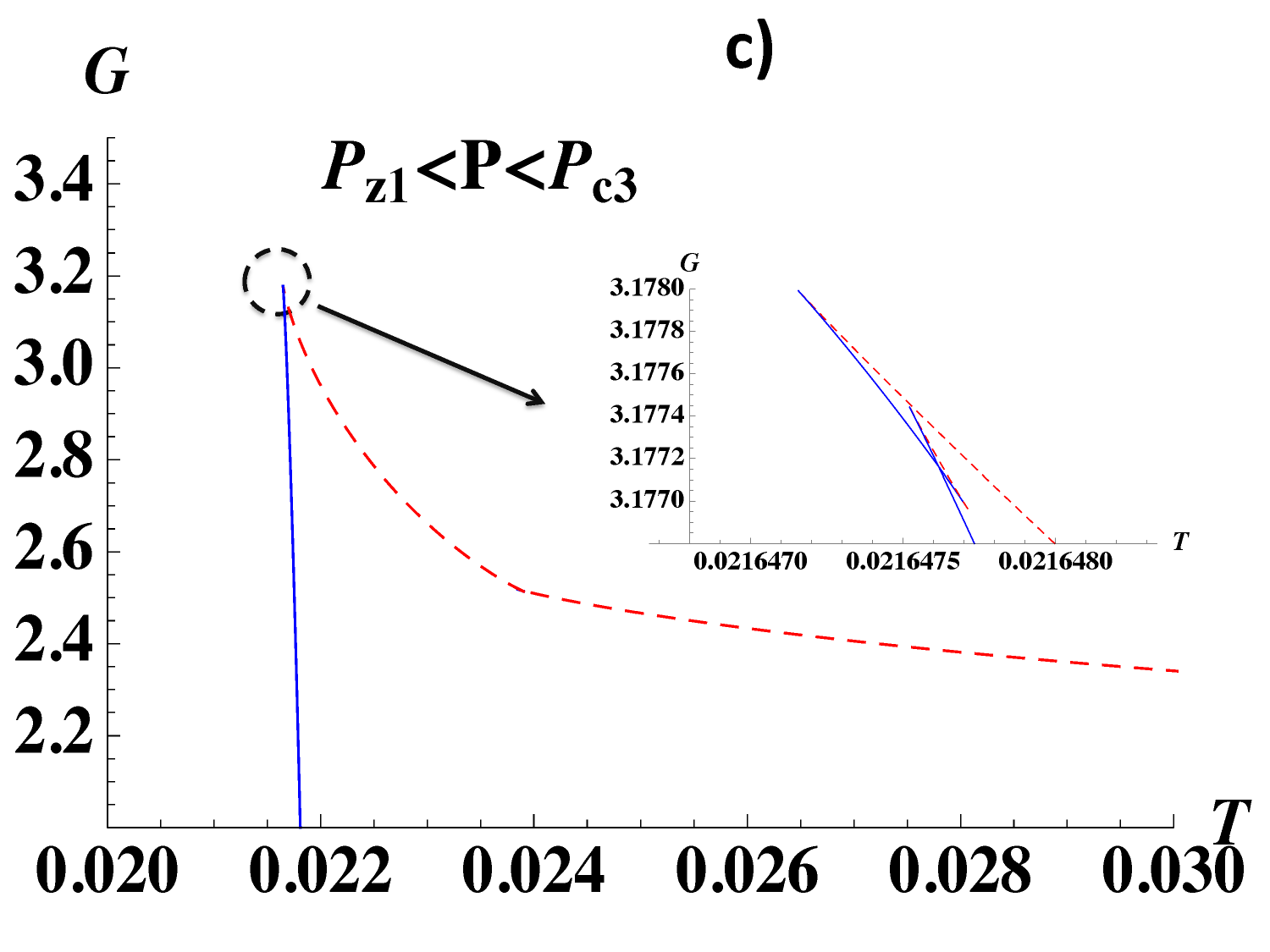}}&
\rotatebox{0}{
\includegraphics[width=0.45\textwidth,height=0.27\textheight]{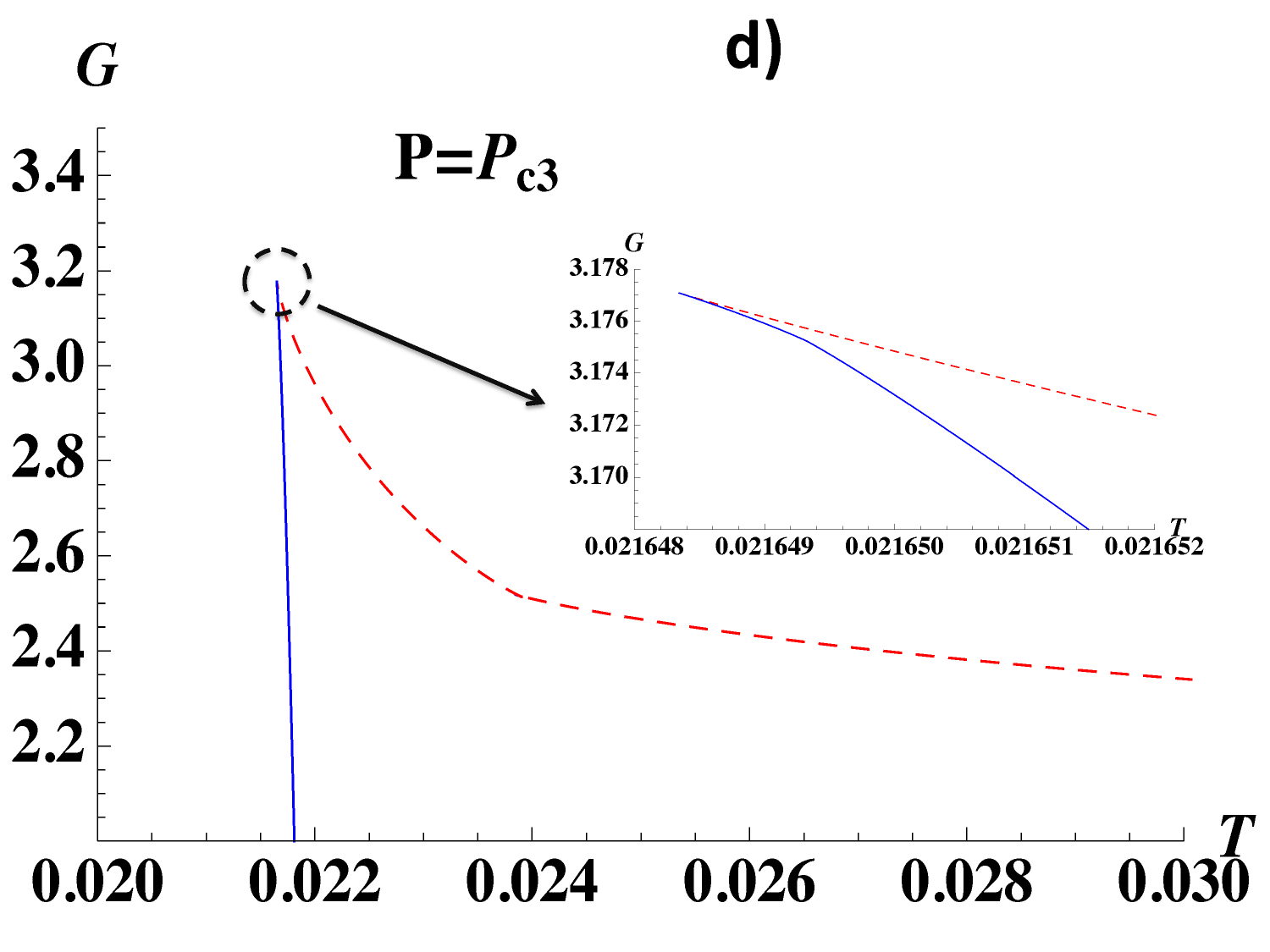}}\\
\rotatebox{0}{
\includegraphics[width=0.45\textwidth,height=0.27\textheight]{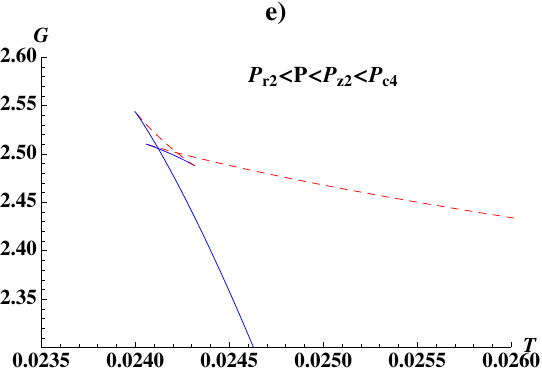}}&
\rotatebox{0}{
\includegraphics[width=0.45\textwidth,height=0.27\textheight]{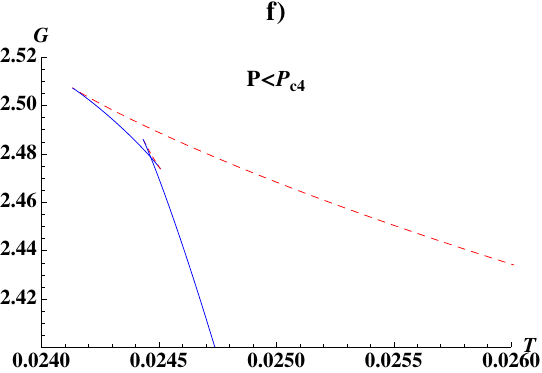}}\\
\\
%width=0.4\textwidth,height=0.33\textheight
\end{tabular}
\caption{ Multiple-RPT for $d=6$, $Q=1$, $\beta=1$ and $\alpha=27$ of GB-BI-AdS black holes. Here, the Gibbs free energy is displayed with respect to temperature for various values of pressures $P=\{0.0007343,0.00073445, 0.0007347,0.00073512,0.0039,0.0051\}$ (from top to bottom). The Solid (blue)/dashed (red) lines correspond to $C_P$ positively and negative, respectively. For $P_{c1}=-0.00126014<P<0$, two zero order phase transitions that minimize the Gibbs free energy occure, similar to Fig. \ref{figure:RE}. In this ranges of pressure, three phases LBH/IBH/SBH are connected by two zero-order phase transitions at $T_1$ and $T_0$, respectively. At $0<P <P_{r1}=0.0007343$, we find a new branch of large thermodynamically stable black holes. At $P=P_{r1}$, the swallowtail which apears between $(P_{c2}=0.000733679,P_{c3}=0.000735122)$ in the unstable branch touches the lower stable branch. We observe one of the RPT of black hole for $P_{r1}=0.0007343< P <P_{z1}=0.00073445$. Since the critical point $P_{c2}$ does not minimize Gibbs free energy, it is an unphysicial critical point. For $P_{z1}<P<P_{c3}$ we have a first order phase transition while disappear at the critical point $P=P_{c3}$. Increasing  pressure, we will have another RPT for $P_{r2}=0.00365 < P <P_{z2}=0.0041$. Above the critical point $P_{c4}=0.00591192$, the system displays no phase transitions.
}\label{figure:RE1}
\end{figure*}

\subsubsection{ Reentrant-Triple points phase transitions}

For $16 \leq \alpha < 18$, we consider Eq. (\ref{critical}) to determine the critical temperature $T_c$ and volume $V_c$ for the specific values of $Q$, $\beta$ and $\alpha$. We replace these critical parameters in the equation of state (\ref{equationstate}) to obtain the critical pressures. Here we find four critical points with the negative and positive pressures of  $P_{c1}$, $P_{c2}$, $P_{c3}$  and  $P_{c4}$. For $\alpha=16.5$, we have one critical point with a negative pressure at $P_{c1}=-0.061$ and three critical points with the positive pressures of  $P_{c2}=0.00115$, $P_{c3}=0.001197$,  and  $P_{c4}=0.00123$. However, only  three of the critical points with the negative pressure of  $P = P_{c1}<0$ and positive pressures of $P = P_{c3}$  and  $P=P_{c4}$  are the physical critical points. The second critical point at $P = P_{c2}$ does not have the minimum Gibss free energy; It is, therefore an unphysical critical point.\\
The resulting Gibbs free energy with respect to temperature for fixed values of $\beta$, $Q$, and $\alpha$ and for  $d = 6$ is illustrated in Fig. \ref{figure:RETR}. \\
At $P = P_{c1}$ with a negative  pressure, we have a second order phase transition in which $C_p$ goes to infinity  (Fig. \ref{figure:RETR}a). For $P_{c1}<P \leq 0$, we observe two zero order phase transitions in Fig. \ref{figure:RETR}b. By increasing  pressure, only one phase of large stable black holes exist for $0 < P \leq P_{r}$ (Fig.  \ref{figure:RETR}c).\\ There are special pressure ranges, $P_{r}=0.00037 \leq P <P_{z}=0.000415$, in which the RPT appears (Fig. \ref{figure:RETR}. d, e). In this range of pressures, the  minimum of Gibss free energy is discontinuous and we have two separate phases of large and small size black holes. These phases are connected by a jump in $G$, or a ‘zeroth-order phase transition’. This critical behavior admits a reentrant large/small/large black hole phase transition.\\
At $P_t=0.001181$, we have two first order phase transitions with equal values of  pressure and temperature similar to those of the triple point in the solid/liquid/gas phase transition (see Fig. \ref{figure:RETR}f). In Fig. \ref{figure:RETR}g, we observe two swallowtails in the stable branch of black holes that minimize $G$ for $P_{c2}<P_{t} \leq P <P_{c3}$. One of these swallowtails starts from $P_{c1}$ and terminates at $P= P_{c3}$. Another swallowtail exists between $P_{c2} \leq P \leq P_{c4}$. Since the critical point at $P=P_{c2}$ does not minimize G, the black holes experience only the critical point at $P=P_{c4}$. \\
Furthermore, in Fig. \ref{figure:RETR}h, we have a critical behavior analogous to Van der Waals system for $P_{c3} \leq P <P_{c4}$, i.e, a first order phase transition between small and large black holes for $P <P_{c4}$ and one critical point at $P =P_{c4}$. Also, the system displays no phase transitions for $P > P_{c4}$ (Fig. \ref{figure:RETR}i).\\
In this specific range of $\alpha$, we observe both reentrant and triple point phenomena in the system for the fixed values of $\beta$, $\alpha$, $Q$ at different pressures.

\subsubsection{ Reentrant phase transitions}

For $18 \leq \alpha < 25$, by using Eqs. (\ref{critical}) and (\ref{equationstate}), the Gibss free energy admits one critical point with a negative pressure of $P_{c1}$ and three critical points with positive pressures of $P_{c2}$, $P_{c3}$  and  $P_{c4}$. For $\alpha=20$, we have one critical point with a negative pressure of $P_{c1}=-0.032$ and three critical points with the positive pressures of $P_{c2}=0.000981$, $P_{c3}=0.0009897$  and  $P_{c4}=0.00197$, two of which $(P=P_{c1} \ and \ P=P_{c4})$ minimize $G$. Also, we observe a large/small/large reentrant phase transition in the specific range of $P_{r}=0.0006 < P < P_{z}=0.00064$.\\
At $P=P_{c1}<0$ we have the thermodynamically unstable branch  with a negative $C_P$ (Fig. \ref{figure:RE}a). For $P_{c1}<P \leq 0$, two zero order phase transitions which minimize the Gibbs free energy at $T_1$ and $T_0$ occur at the stable branch. In this range, the phases LBH/IBH/SBH are connected by two ‘zero-order phase transitions’ at $T_1$ and $T_0$ (Fig. \ref{figure:RE}b).\\
For $0<P <P_r$, we find a new branch of large thermodynamically stable black holes. Thus, only one phase of the large black holes exists.\\
At $P=P_r$, the swallowtail in the unstable branch touches the lower stable branch. We observe the reentrant large/small/large black hole phase transition for $P_{r} < P <P_{z}$ in Fig. \ref{figure:RE}c.\\
Increasing  pressure, we observe two swallowtails. One of them starts from $P=P_{c2}$ and terminates at $P= P_{c3}$. These critical points do not minimize the Gibbs free energy ( Fig. \ref{figure:RE}d). In this situation, the system only sees the swallowtail that occurs at $P=(P_{c1},P_{c4})$. In this situation, the first order phase transition appears at the negative value of pressures $P>P_{c1}$ and terminates at $P<P_{c4}$,  (Fig. \ref{figure:RE} e, f).
Above the critical point $P=P_{c4}$, the system displays no phase transitions. \\

For $13 \leq \alpha < 16$ and $32 \leq \alpha < 40$, we have the reentrant phase transition.
For $13 \leq \alpha < 16$, the Gibss free energy has two critical points with positive pressures, $P=P_{c1}$ and $P=P_{c2}$ for the stable black holes. At $P=P_{c1}>0$, we have the thermodynamically unstable branch  with negative $C_P$ (Fig. \ref{figure:RE0}a). Also, we observe a large/small/large reentrant phase transition in the specific range of $P_{r} < P < P_{z}$ and only one swallowtail occurs in the range of $P_z<P < P_{c2}$ (Fig. \ref{figure:RE0}b, c). At $P=P_{c2}$ we have the second order phase transition (Fig. \ref{figure:RE0}d).\\
For $32 \leq \alpha < 40$, although the black hole system admits two critical points at $P=P_{c1}$ and $P=P_{c2}$ but it is only the second critical point that minimizes the Gibss free energy.\\

\subsubsection{ Multiple-Reentrant phase transitions}

For $25 \leq \alpha < 32$, same as previous calculation we determine the critical temperature $T_c$ and volume $V_c$ for the specific values of $Q$, $\beta$ and $\alpha$ by using Eq. (\ref{critical}). We replace these critical parameters in the equation of state (\ref{equationstate}) to obtain the critical pressuresGibss free energy has four critical points and we observe two reentrant phase transitions. Thus, Gibss free energy has four critical points and we observe two reentrant phase transitions.\\
For the range of $25 \leq \alpha <28$, we have $P_{c1}<0$ with a thermodynamically unstable branch or a negative $C_P$. For $P_{c1}<P\leq 0$, two zero order phase transitions occur which minimize the Gibbs free energy at $T_1$ and $T_0$. In this range of pressures, the phases LBH/IBH/SBH are connected by two ‘zero-order phase transitions’ at $T_1$ and $T_0$, respectively.\\
For $0<P <P_{r1}$, we find a new branch of large thermodynamically stable black holes. In Fig. \ref{figure:RE1}a, at $P_{c2}<P=P_{r1}<P_{c3}$, the swallowtail in the unstable branch touches the left stable branch. Thus, we observe the reentrant large/small/large black hole phase transition for $P_{r1} < P <P_{z1}$ (Fig. \ref{figure:RE1}b). Since the critical point $P_{c2}$ where this reentrant phase transition appears does not minimize Gibbs free energy, so this critical point is unphysicial (Fig. \ref{figure:RE1}, c, d). By increasing  pressure, this swallowtail disappears and we have another reentrant phase transition for $P_{r2} < P <P_{z2}$ (Fig. \ref{figure:RE1} e, f). In this situation, three of the critical points are  physical.\\
For $28 \leq \alpha <32$, we have the same critical behavior (two reentrant phase transitions) but the critical point at $P=P_{c1}$ has a positive pressure. In this condition, a new branch of large thermodynamically stable black holes appears at $P=P_{c1}$ and  the two ‘zero-order phase transitions’ at $T_1$ and $T_0$ do not minimize the Gibbs free energy. Thus, only one phase of the large black hole exists at $P=P_{c1}$ and both the critical points at $P=P_{c1}$ and $P=P_{c2}$ are unphysical. Similar to the previous case (for $25 \leq \alpha <28$), we observe two reentrant phase transitions for $P_{r1} < P <P_{z1}$ and for $P_{r2} < P <P_{z2}$ (Fig. \ref{figure:RE1}). The results are summarized in Table. \ref{canonical}.
\begin{figure}
\begin{center}
  \includegraphics[width=8.5cm,height=6cm]{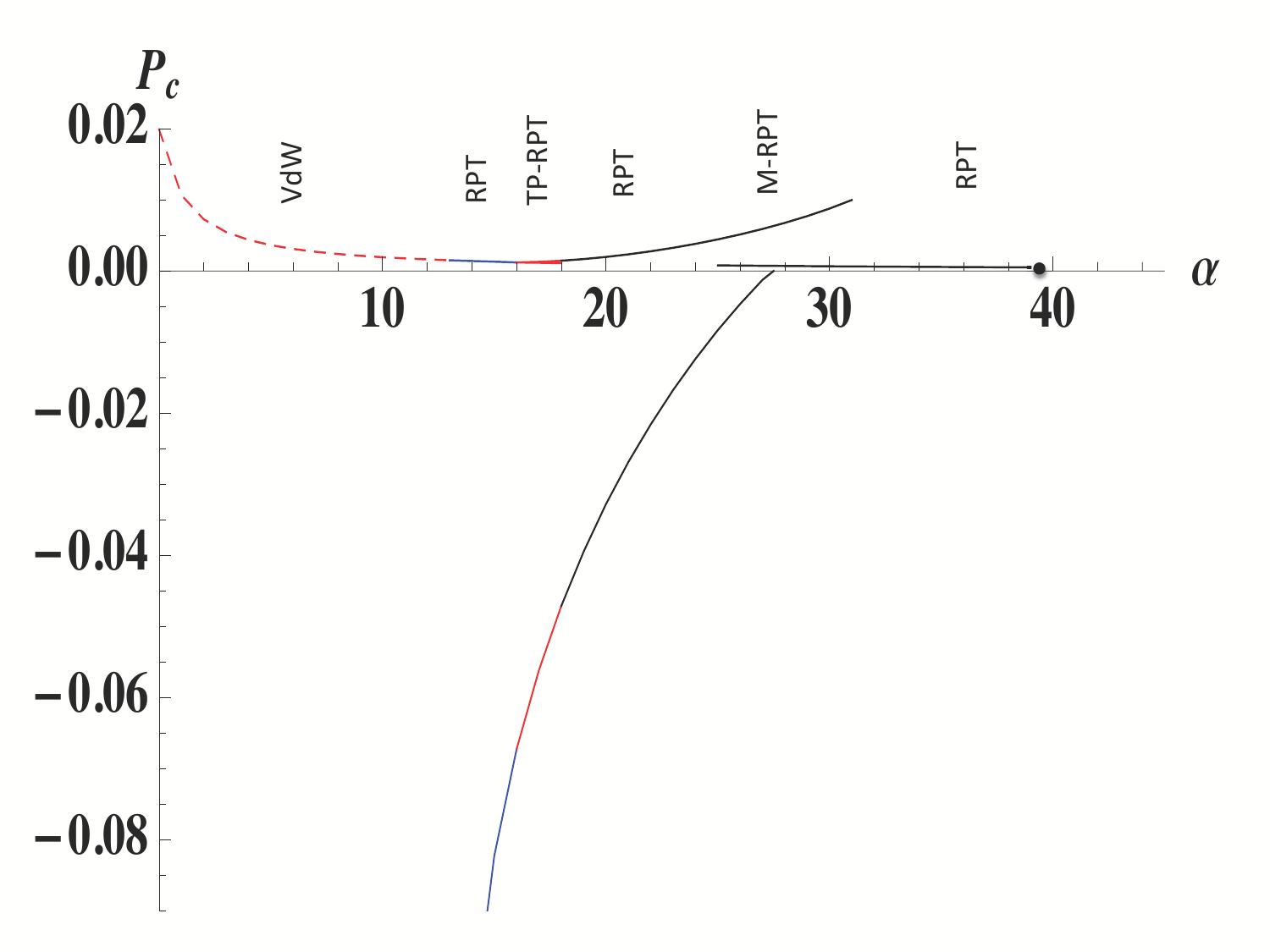}\\
  \caption {\it{The critical values of pressures with respect to $\alpha$ for $\beta=1$ and $Q=1$ of GB-BI-AdS black holes. For $0\leq \alpha<16$, there is one critical point and the system behaves similar to a "standard liquid/gas". We have a minimum pressure at $\alpha=16$ in which  both the reentrant and triple point behaviors appear.}}\label{figure:cirit}
\end{center}
 \end{figure}

The critical values of pressure with respect to $\alpha$ for $\beta=1$ and $Q=1$ are presented in Fig. \ref{figure:cirit}. Based on the critical behaviors, we can divide the diagram to different regions. For $0\leq \alpha<13$,  one critical point exists, the system behaves similar to a "standard liquid/gas", and we observe RPT for $13\leq \alpha<16$. We have a minimum critical pressure at $\alpha=16$ in which  both reentrant and triple point behaviors appear.  For $16 \leq \alpha < 18$ we observe RPT and TP with increasing pressure.\\
 For $18 \leq \alpha < 25$, we observe four critical points  two of which minimize the Gibbs free energy. Hence, the GB-BI-AdS black holes experience a reentrant phase transition while we have not triple point phase transitions in this range. In the case of $25 \leq \alpha <32$, the GB-BI-AdS black holes have multiple-RPT in different pressure ranges.\\
Also, one critical point occure for the range  $32 \leq \alpha < 40$. The system has a reentrant phase transition for the specific range of pressure (Fig. \ref{figure:RE0}). For $\alpha \geq 40$, we have two critical points, but neither minimizes the Gibbs free energy and they are unphysical critical points. Thus, above $\alpha=40$, only one phase of large black holes exist and there is no phase transition (Fig. \ref{figure:cirit}).\\
We expand our calculation when two parameters $\beta$ and $Q$ are small enough ($\beta=0.01$, $Q=1$ or $\beta=1$, $Q=0.01$), the results show that we have only a reentrant phase transition for $\alpha \neq 0$. Thus, the black hole can not see those different critical behaviours by increasing $\alpha$ for the small ranges of $\beta$ and $Q$. Also, for large values of $Q=100$,  $\beta=1$ and $\alpha\neq 0$ they tend to the VdW behaviour. For large values the BI parameter $\beta=1000$ when $Q=1$ and $\alpha \neq 0$, the black holes experience only the triple point.

\begin{figure}
\begin{center}
  \includegraphics[width=8.5cm,height=6.8cm]{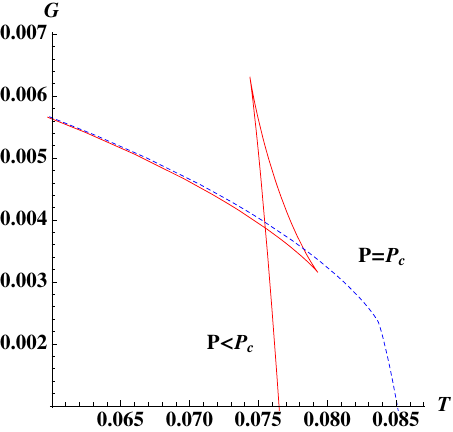}\\
  \caption {\it{The Gibbs free energy with respect to $T$ for $\beta=1$, $\alpha=\frac{16}{100}$ and $\phi=\frac{6}{10}$ of GB-BI-AdS black holes for $d=5$ in the  grand canonical ensemble. We have one critical point at $P_c=0.021$ where the Van der Waals behavior occurs.}}\label{figure:REgrand}
\end{center}
 \end{figure}
%%%%%%%%%%%%%%%%%%%%%%%%%%%%%%%%%%%%%%%%%

\subsection{Critical behavior in the grand Canonical ensemble }

The phase transitions and critical exponents of the  AdS-GB black holes for $d$ dimensions and in the grand canonical ensemble are considered in \cite{GBgrand}. The phase transitions of BI-AdS black holes in $d=4$ in the grand canonical ensemble were recently studied in \cite{Wei:2012ui}.
Here, we investigate the phase transitions of  GB-BI-AdS black holes in the grand canonical ensemble while $\beta$, $\phi$, and $\alpha$ are thermodynamic variables. \\
Let us define a variable $x$ for $d=5$ as,
\begin{eqnarray}
\label{definition}
&&x=\frac{256 \pi Q}{(3 v) ^{3}} \sqrt{\frac{1}{3}},\\
&&v=\frac{8 \phi }{3 \sqrt{3} x \, _2F_1[\frac{1}{3},\frac{1}{2},\frac{4}{3},-\frac{3 x^2}{\beta ^2}]},
\end{eqnarray}
where, $v$ is the specific volume. By using Eqs. (\ref{potential}) and (\ref{equationstate}) and the above definition, we can rewrite the equation of state in terms of the electric potentials $\phi$ and $x$ for $d=5$ in the following form:
\begin{eqnarray}
\label{equationstategrand}
&&P=\frac{1}{32 \sqrt{3} \pi  \phi ^3} \Bigr(36 \pi  T x \phi ^2 \, _2F_1[\frac{1}{3},\frac{1}{2},\frac{4}{3},-\frac{3 x^2}{\beta ^2}]\\\nonumber
&&+54 \pi  \alpha  T x^3 \  _2F_1[\frac{1}{3},\frac{1}{2},\frac{4}{3},-\frac{3 x^2}{\beta ^2}]^3-8 \sqrt{3} \beta ^2 \phi ^3\\\nonumber
&&-9 \sqrt{3} x^2 \phi  \  _2F_1[\frac{1}{3},\frac{1}{2},\frac{4}{3},-\frac{3 x^2}{\beta ^2}]^2+8 \sqrt{3} \beta ^2 \phi ^3 \sqrt{\frac{\beta ^2+3 x^2}{\beta ^2}} \Bigl),
\end{eqnarray}
The critical points can be determined by using the conditions
\begin{eqnarray}
\frac{\partial P}{\partial v}=0, \ \ \frac{\partial ^2 P}{\partial v ^2}=0
\end{eqnarray}

\begin{figure*}
\centering
\begin{tabular}{ccc}
\rotatebox{0}{
\includegraphics[width=0.3\textwidth,height=0.22\textheight]{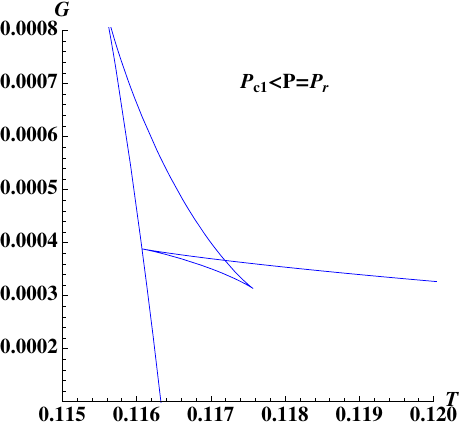}}&
\rotatebox{0}{
\includegraphics[width=0.3\textwidth,height=0.22\textheight]{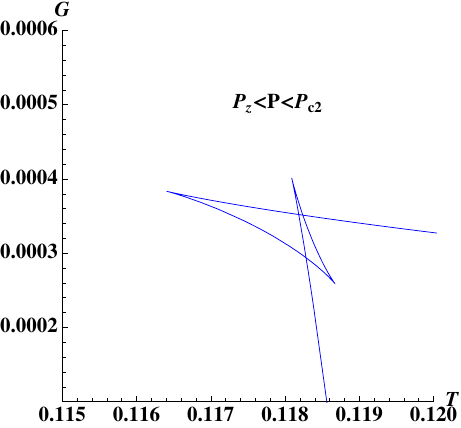}}&
\rotatebox{0}{
\includegraphics[width=0.3\textwidth,height=0.22\textheight]{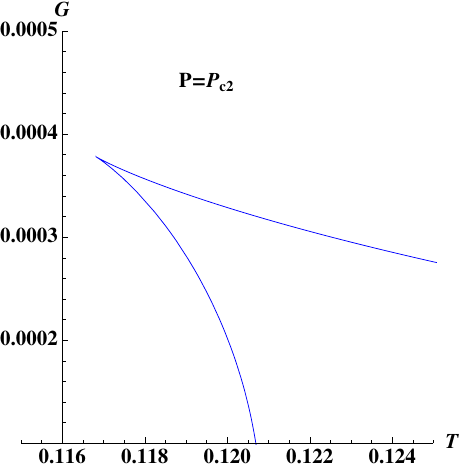}}\\
\\
%width=0.4\textwidth,height=0.33\textheight
\end{tabular}
  \caption {\it{The Gibbs free energy with respect to $T$ for $\beta=1$, $\alpha=\frac{16}{100}$ and $\phi=\frac{6}{10}$ of GB-BI-AdS black holes for $d=6$. We have two critical points with positive pressures in which the reentrant phase transition occurs and disappears.}}\label{figure:REgrand1}
\end{figure*}

The results show that there is only one critical point that depends on the coupling coefficients $\alpha$, $\beta$, and $\phi$ for $d=5$. \\
The Gibbs free energy in the grand canonical ensemble is given by
\begin{eqnarray}
G(x,\phi,\alpha,\beta)=M-T S-Q \phi .
\end{eqnarray}
By using the above definitions and Eqs. (\ref{massGB}), (\ref{entropy}), and  (\ref{definition}), the Gibss free energy with respect to temperature is displayed for $d=5$ in Fig. \ref{figure:REgrand}. The GB-BI-AdS black holes behave similar to the Van der Waals fluid for $d=5$ for all the values of $\beta$, $\alpha$, and $\phi$.\\
Let us expand these calculations to the case with  $d = 6$. In this case, we observe a reentrant phase transition with two critical points for the specific ranges of  $\frac{6}{100} \leq \alpha \leq \frac{18}{100}$ and $\phi=\frac{6}{10}$ in $d=6$  (Fig. \ref{figure:REgrand1}) while for $\frac{6}{100} \geq \alpha$,  $\alpha \geq \frac{18}{100}$ and $\phi=\frac{6}{10}$ we have the Van der Waals behavior. The same critical behaviours are observed for different ranges (small and large values) of $\beta$ and $\phi$, it means that GB-BI-AdS black holes in grand canonical ensembles in $d=6$ behave critically the same for various values of the BI parameter $\beta$ and $\alpha$.\\
We also observe the Van der Waals behavior in the BI-AdS black holes, ($\alpha=0$), in the grand canonical ensemble and for $d =5,6$. The diagram is similar to  Fig. \ref{figure:REgrand}.\\
Now, we consider the limit of $\beta \to \infty$ for the charged-GB-AdS  black holes. It is shown that these black holes in the grand canonical ensemble and for $d=5$ behave similar to the standard liquid/gas of the Van der Waals fluid \cite{GBgrand}.\\
Also, one phase of the large black holes for $d \geq 6$ of charged-GB-AdS black holes  was investigated in the grand canonical ensemble \cite{Mirza:2014}.\\

%%%%%%%%%%%%%%%%%%%%%%%%%%%%%%%%%%%%%%%%%%%%%%%%%%%%%%%%%%%%
\section{AdS-Born-Infeld black holes in the third order lovelock gravity}

\begin{table*}
\caption{Critical behaviors of  black holes for $Q=1$ and $\beta=1$ in the canonical ensemble}
\label{canonical}
\begin{tabular*}{\textwidth}{@{\extracolsep{\fill}}lrrrrl@{}}
\hline
Black holes system&d& &Critical behaviors &     &  \\
\hline
$GB-BI-AdS$& 5  & &$Van \ der \ Waals \ (VdW)$& &  \\
\hline
$GB-BI-AdS$& 6    & $0 \leq \alpha<13$&  $13 \leq\alpha<16$, $18 \leq \alpha<25$ and $32 \leq \alpha<40$  \\
 &  & $VdW$\  \   \    &$RPT$ \  \  \  \  \  \   \  \  \  \  \   \    \  \  \  \  \     \\
&     & $16 \leq \alpha<18$&$25 \leq \alpha<32$ \  \  \   \    \  \     \  \  \   \   \  \  \   \    \  \     \  \  \   \    \  \   \  \  \\
 &   &$RPT-TP$&$multiple-RPT $   \  \  \   \    \  \     \  \  \   \    \  \  \  \   \    \  \  \\
\hline
$BI-AdS$& 5,6  &  $VdW$&& \\
\hline
\end{tabular*}
\end{table*}

The action of the third order Lovelock gravity with a nonlinear Born-Infeld
 electromagnetic field in the $d$-dimensional space time is given by \cite{Dehghani2008}
\begin{eqnarray}
I=\frac{1}{16\pi }\int d^{n+1} x\sqrt{-g}(R-2\Lambda+\alpha_2 {\cal{L}}_2+\alpha_3 {\cal{L}}_3+{\cal{L_F}}),\nonumber
\end{eqnarray}
where,
 \begin{eqnarray}
{\cal{L}}_2=R_{\mu \nu \gamma \delta} R^{\mu \nu \gamma \delta }-4 R_{\mu \nu} R^{\mu \nu }+R^{2},
\end{eqnarray}

\begin{eqnarray}
\nonumber & {\cal{L}}_3={{R}^{3}}+2{{R}^{\mu \nu \sigma \kappa }}{{R}_{\sigma \kappa \rho \tau }}R_{\mu \nu }^{\rho \tau }+8R_{\sigma \rho }^{\mu \nu }R_{\nu \tau }^{\sigma \kappa }R_{\mu \kappa }^{\rho \tau }\\
\nonumber &+24{{R}^{\mu \nu \sigma \kappa }}{{R}_{\sigma \kappa \nu \rho }}R_{\mu }^{\rho }+3R{{R}^{\mu \nu \sigma \kappa }}{{  R}_{\sigma \kappa \mu \nu }}\\
\nonumber &+24{{R}^{\mu \nu \sigma \kappa }}{{R}_{\sigma \mu }}{{R}_{\kappa \nu }}+16{{R}^{\mu \nu }}{{R}_{\nu \sigma }}R_{\mu }^{\sigma }-12R{{R}^{\mu \nu }}{{R}_{\mu \nu }},\\
\end{eqnarray}
and
 \begin{eqnarray}
 {\cal{L_F}}=4{{\beta }^{2}}(1-\sqrt{1+\frac{F^{\mu \nu} F_{\mu \nu }}{2{{\beta }^{2}}}}),
\end{eqnarray}
 $F_{\mu \nu}$ is the vector potential.

\begin{figure*}
\centering
\begin{tabular}{cc}
\rotatebox{0}{
\includegraphics[width=0.5\textwidth,height=0.27\textheight]{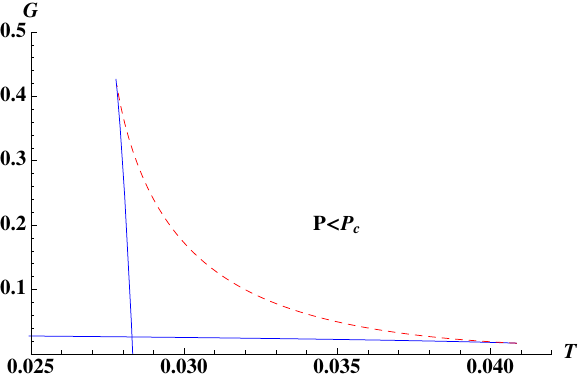}}&
\rotatebox{0}{
\includegraphics[width=0.5\textwidth,height=0.27\textheight]{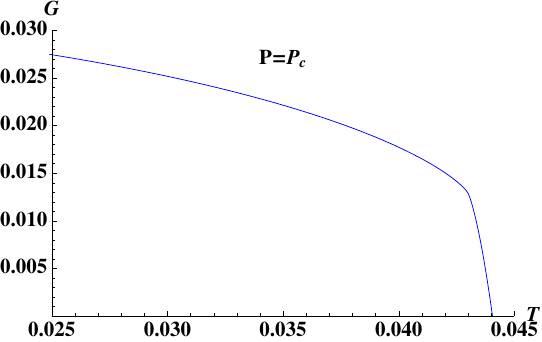}}\\
\\
%width=0.4\textwidth,height=0.33\textheight
\end{tabular}
\caption{ \it{The Gibbs free energy  with respect to $T$ of the third order of Lovelock-BI-AdS black holes for $d=7$, $\phi=7/10$, $\beta=1$ and $\alpha=1$. Here, dashed red and solid blue lines
correspond to positive, and negative $C_{P}$, respectively. At $P<P_c$, the black hole experiences a first order phase transition and at $P=P_c$ we have a second order phase transition. We do not have a phase transition at $P>P_c$.}
}\label{figure:Lovgrand}
\end{figure*}

\noindent In the above action, $\beta $, ${{\alpha }_{2}}$, and ${{\alpha }_{3}}$ are the Born-Infeld parameters, the second and third order Lovelock coefficients, respectively.
Let us consider the following case
 \begin{eqnarray}
 \alpha _2=\frac{\alpha }{(d-3)(d-4)},
\end{eqnarray}
 \begin{eqnarray}
\alpha _3=\frac{\alpha^2}{72(
\begin{matrix}
  d-3\\
  4\\
\end{matrix})}
\end{eqnarray}
The metric of the $d$-dimensional static solution is as follows:
\begin{eqnarray}
 \ d{{s}^{2}}=-f(r)d{{t}^{2}}+\frac{d{{r}^{2}}}{f(r)}+r^2 h_{ij} dx ^i dx^j,
\end{eqnarray}
Here, $h_{ij}$ denotes the line element of $(d-2)$-dimensional hypersurface. Setting $k=1$, we have
 \begin{eqnarray}
 f(r)=1+\frac{{{r}^{2}}}{\alpha }(1-g{{(r)}^{1/3}}),
\end{eqnarray}
where,
\begin{eqnarray}
\label{gr}
g(r)&=&1+\frac{3\alpha m}{r^{d-1}}-\frac{12\alpha {{\beta }^{2}}}{(d-1)(d-2)} \Bigr(1\\\nonumber
&-&\sqrt{1+\frac{(d-2)(d-3){{q}^{2}}}{2{{\beta }^{2}}{{r}^{2d-4}}}} \Bigl)+\frac{8 \pi P }{2{{\beta }^{2}}}\\\nonumber
&+&\frac{(d-2)}{(d-3)} \frac{(d-2)(d-3) q^{2}}{2{{\beta }^{2}}{{r}^{2d-4}}} \ {}_2 F(r), \\
\end{eqnarray}
and ${}_{2}F(r )$ is the hypergeometric function
  \begin{eqnarray}
 \label{29b}
{}_{2}F(r )={}_{2} F_1 [\frac{1}{2},\frac{d-3}{2d-4},\frac{3d-7}{2d-4},-\frac{(d-2)(d-3){{q}^{2}}}{2{{\beta }^{2}}{{r}^{2d-4}}}],
\end{eqnarray}
where,
\begin{eqnarray}
Q=\frac{q \sum _{d-2}}{4 \pi} \sqrt{\frac{(d-2)(d-3)}{2}}.
\end{eqnarray}
The ADM mass of the black holes is
\begin{eqnarray}
 \label{masslove}
M&=&\frac{(d-2){{\Sigma }_{d-2}}}{16\pi }m\\\nonumber
&=&\frac{(d-2){{\Sigma }_{d-2}}r_{+}^{d-1}}{48\pi \alpha} \Bigr(\frac{{{(r_{+}^{2}+\alpha )}^{3}}}{r_{+}^{6}}-1+\frac{12\alpha {{\beta }^{2}}}{(d-1)(d-2)}\\\nonumber
&\times& (1+\frac{8 \pi P }{2{{\beta }^{2}}}-\sqrt{1+\eta_+ }+\frac{(d-2) {}_{2}F(r_+ )\eta_+ }{d-3})\Bigl),
\end{eqnarray}
where, ${{\Sigma }_{d-2}}$ denotes the volume of $(d-2)$-dimensional hypersurface and we set ${{\Sigma }_{d-2}}=1$ for simplicity. Also, $r_+$ is calculated from $f(r_+)=0$.
The thermodynamic quantities are given by
\begin{eqnarray}
 \label{templove}
T&=&\frac{1}{12(d-2){{(r_+^{2}+\alpha )}^{2}} {{r}_{+}}} \Bigr(12r_{+}^{6}\beta ^2+6\pi P r_{+}^{6} \\\nonumber
&-&12r_{+}^{6}\beta ^2 \sqrt{1+{{\eta_+}}}+(d-2) \ \Bigr(3(d-3)r_+^{4}\\\nonumber
&+&3(d-5)\alpha r_{+}^{2}+(d-7){{\alpha }^{2}} \Bigl)\Bigl),\\
\label{entropylove}
S&=&\frac{(d-2)r_{+}^{d-6}}{4}(\frac{r_{+}^{4}}{d-2}+\frac{2r_{+}^{2}\alpha }{d-4}+\frac{{{\alpha }^{2}}}{d-6}),\\
\phi&=& \frac{4 \pi Q }{(d-3) r_+^{d-3}} {}_{2}F(r_+),
 \end{eqnarray}
Here, $\eta_+=\frac{16 \pi^2 Q^{2}}{{{\beta }^{2}}{{r}^{2d-4}_+}}$ and the above thermodynamic quantities are valid for $d \geq 7$.\\
The pressure is proportional to the cosmological constant. We identify the pressure of the black hole in the extended phase space with the following form
 \begin{eqnarray}
 \ P=-\frac{\Lambda }{8\pi }.
 \end{eqnarray}
Since the cosmological constant is considered as the thermodynamic pressure, we replace the ADM mass of the black hole by the enthalpy. So, the first law of thermodynamics and its relevant quantities read as follow
\begin{eqnarray}
 \label{33b}
 \ dM=TdS+\phi dQ+VdP+{\cal{A}}d\alpha +{\cal{B}} d\beta .
\end{eqnarray}
The parameteres $V$, $\cal{A}$, and $\cal{B}$ are the thermodynamic quantities conjugating to the pressure $P$, the Gauss Bonnet coefficient $ \alpha$, and the Born Infeld parameter $\beta$, respectively
\begin{eqnarray}
V&=&\frac{r_+^{d-1}}{d-1}=\frac{1}{d-1}(\frac{(d-2) \it{v}}{4})^{d-1},\\
{\cal{A}}&=&\frac{(d-2)r_+^{d-7}\left(3r_+^2+2\alpha \right)}{48\pi }\\\nonumber
&-&\frac{1}{2}(d-2)r_+^{d-6}T\left(\frac{r_+^2}{d-4}+\frac{\alpha }{d-6}\right),\\
 \label{35b}
{\cal{B}}&=&\frac{r_+^{d-1}}{2  \pi  \beta  (d-1)} \Bigl(\beta ^2 -\beta ^2  \sqrt{\frac{16 \pi^2 Q^2 r_+^{4-2 d}}{\beta ^2}+1}\\\nonumber
&+&\frac{8 \pi^2 Q^2 r_+^2}{r_+^{2 d-2} } \, _2F_1[\frac{1}{2},\frac{d-3}{2 d-4},\frac{7-3 d}{4-2 d},-\frac{16 \pi^2 Q^2 r_+^{4-2 d}}{\beta ^2}]
\Bigr).
 \end{eqnarray}
In next section, we may assume the critical behaviors and phase transitions of  BI-AdS black holes in the third order of Lovelock gravity and in the grand canonical ensemble.
\begin{figure*}
\centering
\begin{tabular}{ccc}
\rotatebox{0}{
\includegraphics[width=0.3\textwidth,height=0.22\textheight]{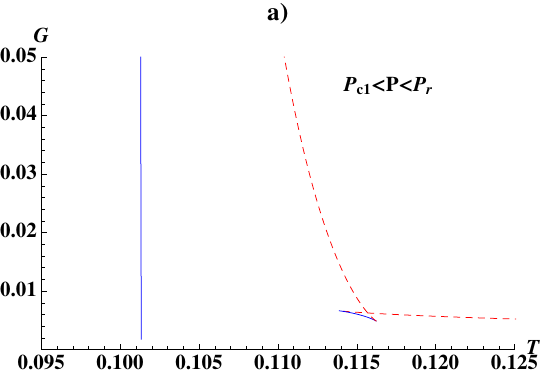}}&
\rotatebox{0}{
\includegraphics[width=0.3\textwidth,height=0.22\textheight]{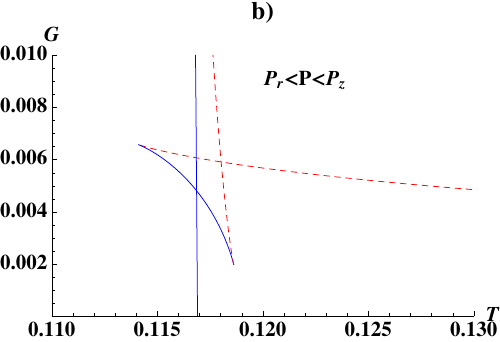}}&
\rotatebox{0}{
\includegraphics[width=0.3\textwidth,height=0.22\textheight]{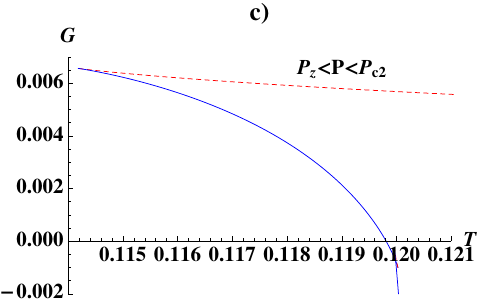}}\\
\\
%width=0.4\textwidth,height=0.33\textheight
\end{tabular}
\caption{ \it{Reentrant phase transition for $d=8$, $\phi=0.5$, $\beta=1$ and $\alpha=1$ of the third order of Lovelock-BI-AdS black holes. Here, the Gibbs free energy is displayed with respect to temperature for various values of pressure, with pressure increasing from right to left. The solid-blue/dashed-red lines correspond to positive and negative  $C_P$, respectively. At $P=P_{c1}$, we have the thermodynamically unstable branch. At $P <P_{r}$ appears a new branch of large thermodynamically stable black holes. Also, we observe the reentrant large/small/intermediate black hole phase transition for $P_{r} \leq P <P_{z}$. For $P_{z}<P<P_{c2}$, we have a first order phase transition with  negative values of Gibbs free energy. This first order phase transition disappears and we have a second order phase transition at the critical point $P=P_{c2}$.}
}\label{figure:LovgrandRE}
\end{figure*}

%%%%%%%%%%%%%%%%%%%%%%%%%%%%%%%%%%%%%%%%%%%%%%%%%%%

\subsection{Critical behavior in the grand Canonical ensemble}

The phase transitions and critical behaviors of  charged-AdS balck holes  in the third order of Lovelock gravity have been  investigated in the canonical ensemble \cite{Hao:2014,Mann:2014}.  Also, the third order Lovelock-BI-AdS black holes have been studied in the canonical ensemble and  $d=7$ in \cite{MoLiu:2014}.\\
Now, let us assume the third order Lovelock-BI-AdS black holes in the grand Canonical ensemble in which the electric potentials $\phi$, $\alpha$, and $\beta$ are considered as thermodynamic variables.\\
Similar to our previous calculations, we define a new parameter $x$ for these black holes for $d=7$,  as in the following:
\begin{eqnarray}
\label{definitionlove}
&&x=\frac{2048 \sqrt{\frac{2}{5}} \pi  Q}{3125 v^5},\\
&&v=\frac{8 \sqrt{\frac{2}{5}} \phi }{5 x \, _2F_1[\frac{2}{5},\frac{1}{2},\frac{7}{5},-\frac{10 x^2}{\beta ^2}]},
\end{eqnarray}
Here, $v$ is the specific volume. By using Eq. (\ref{templove}), the equation of state for $d=7$ and the grand Canonical ensemble for the third order Lovelock-BI in the AdS space are given by
\begin{eqnarray}
P&=&\frac{1}{512 \sqrt{10} \pi  \phi ^5} (1600 \pi  T x \phi ^4 \, _2F_1(r_+)\\\nonumber
&+&625 \pi  \alpha ^2 T x^5 \, _2F_1(r_+)^5+2000 \pi  \alpha T x^3 \phi ^2 \, _2F_1(r_+)^3\\\nonumber
&-&400 \sqrt{10} x^2 \phi ^3 \, _2F_1(r_+)^2-125 \sqrt{10} \alpha  x^4 \phi  \, _2F_1(r_+)^4\\\nonumber
&-&128 \sqrt{10} \beta ^2 \phi ^5+128 \sqrt{10} \beta ^2 \phi ^5 \sqrt{\frac{\beta ^2+10 x^2}{\beta ^2}}).
\end{eqnarray}
The critical points can be determined by using the conditions
\begin{eqnarray}
\frac{\partial P}{\partial v}=0, \ \ \frac{\partial ^2 P}{\partial v ^2}=0
\end{eqnarray}
The Gibbs free energy is given by
  \begin{eqnarray}
 G=M-T S-Q \phi.
 \end{eqnarray}
One is able to obtain the above Gibbs free energy by using Eqs. (\ref{masslove}),  (\ref{templove}), (\ref{entropylove}), and  (\ref{definitionlove}).
We plot the diagram of Gibbs free energy with respect to temperature for these black holes for $d=7$ in Fig. \ref{figure:Lovgrand}. The results show that the system has the Van der Waals behavior for all given values of the parameters $\alpha$, $\beta$, and $P$ for $d=7$ (Fig. \ref{figure:Lovgrand}).\\
For $\phi > 0.42$, we have one critical point with a positive value of the Gibbs free energy.  By decreasing $\phi$, we observe that one critical point for $\phi< 0.42$ occurs with the negative values of the Gibbs free energy which is more stable compare to pure AdS space.  \\
We expand our calculations to the third order Lovelock-BI-AdS black holes for $d=8$. We observe a reentrant phase transition for $\beta=1$, $\alpha=1$, and $\phi \leq 0.6$. This RPT appears between the critical points $P_{c1}$ and $P_{c2}$ in Fig. \ref{figure:LovgrandRE}b. At $P=P_{c2}$, the second critical point, the Gibbs free energy has  negative values (Fig. \ref{figure:LovgrandRE}c). \\
The third order Lovelock-BI-AdS black holes has only one phase of the large black holes for $\phi > 0.6$, $\beta=1$ and $\alpha=1$. \\
The same critical behaviours we obtain for different ranges of $\beta$ and $\phi$. It means that the critical behaviours of third order Lovelock-BI-AdS black holes in grand canonical ensembles in $d=8$ is similar for all values of the parameter $\beta$.\\
Although  the critical values of thermodynamic quantities of these black holes depend on the BI parameter $\beta$ and the coupling coefficient $\alpha$, the critical behaviors and the types of phase transitions depend only on the coupling coefficient $\alpha$ and $\phi$.\\
 In the next subsection, we consider the limit of $\beta \to \infty $ for the charged AdS black holes in the third order of Lovelock gravity.

\begin{table*}
\caption{Critical behaviors of  black holes for $\beta=1$ in the grand canonical ensemble}
\label{grandcanonic}
\begin{tabular*}{\textwidth}{@{\extracolsep{\fill}}lrrrrl@{}}
\hline
Black holes system&d&Critical behaviors &      \\
\hline
$GB-BI-AdS$& 5  &$VdW$\   \  \  \  &   \\
\hline
$GB-BI-AdS$& 6    & $\alpha<\frac{6}{100}$, \  $\frac{18}{100}<\alpha$, \ $\phi=0.6$    \\
 &  & $VdW$   \  \   \  \    \  \   \  \ \\
&     & $\frac{6}{100} \leq \alpha \leq \frac{18}{100}$ , \ $\phi=0.6$  \\
 &  & $RPT$  \  \   \  \   \  \   \  \  \\
\hline
$BI-AdS$& 5,6      &  \    \   $For \ all \ values \ of \ \phi$ \\
&  &$VdW$  \   \   \  \  \    \\
\hline
$Third \ order \ lovelock-BI-AdS$& 7 &$For \ all \ values \ of \ \phi$\\
 &  & $VdW$ \   \  \  \     \  \\
\hline
$Third \ order \ lovelock-BI-AdS$& 8 &$\phi \leq 0.6$ ,\ $\alpha=1$ \\
 &  & $RPT$\  \   \    \   \  \  \\
\hline
$Charged-lovelock-AdS$& 7 &$For \ all \ values \ of \ \phi$  &    \\
&  &$VdW$  \   \  \  \  \ \\
\hline
$Charged-lovelock-AdS$& 8 &$\phi \leq 0.58$ ,\ $\alpha=1$&   \\
 &  & $RPT$\  \   \  \  \   \  \  \\
\hline
\end{tabular*}
\end{table*}

%%%%%%%%%%%%%%%%%%%%%%%%%%%%%%%%%%%%%%%%%%%%%%%%%
\subsection{The charged-AdS third order Lovelock black holes in the grand canonical ensemble}

Let us consider the limit of $\beta \to \infty $ to concentrate on the charged black holes in  the third order of Lovelock gravity. In this case, i.e. $\beta \to \infty $, the Born-Infeld Lagrangian reduces to the Maxwell form and ${}_2 F(r_+ )\to 1$ in Eq. (\ref{29b}). Thus,  Eq. (\ref{gr}) reduces to the following form
 \begin{eqnarray}
 \ g(r)\to 1+\frac{3\alpha m}{{{r}^{d-1}}}+\frac{6\alpha \Lambda }{(d-1)(d-2)}-\frac{3\alpha {{Q}^{2}}}{{{r}^{2d-4}}}.
 \end{eqnarray}
Now, we define a parameter $x$ in the form
\begin{eqnarray}
\label{definitionlove1}
&&x=4 \sqrt{2} \pi  \sqrt{\frac{1}{(d-3) (d-2)}} Q r^{2-d},\\\nonumber
&& r_+=\frac{\sqrt{\frac{2 (d-1)-4}{d-2}} \phi }{x \, _2F_1[\frac{1}{2},\frac{d-3}{2 (d-1)-2},\frac{3 (d-1)-4}{2 (d-1)-2},-\frac{(d-3) (d-2) x^2}{2 \beta ^2}]}.
 \end{eqnarray}
Using Eqs. (\ref{masslove}), (\ref{templove}), and (\ref{definitionlove1}), the Mass and Hawking temperature of the charged-AdS third order Lovelock black holes in the grand Canonical ensemble turn into the following forms
  \begin{eqnarray}
\label{definitionmasslove}
M&=&\frac{2^{\frac{d-15}{2}} \sqrt{d-2} \ x \left(\sqrt{\frac{d-3}{d-2}} \frac{\phi }{x}\right)^d}{3 \pi  (d-3)^{7/2} (d-1) \phi ^7} \Bigl(6 \alpha  (d-2)^3\\\nonumber
&&\left( d^2-4 d+3\right) x^4 \phi ^2+12 (d-3)^2 \left(d^2-3 d+2\right) \\\nonumber
&&\left(2 d \phi ^2+d-6 \phi ^2-2\right) x^2 \phi ^4+384 \pi  (d-3)^3 P \phi ^6\\\nonumber
&&+\alpha ^2 (d-2)^4 (d-1) x^6\Bigr),\\
\label{definitiontemlove}
T&=&\frac{x}{12 \pi  \sqrt{\frac{2 (d-1)-4}{d-2}} (d-2) \phi  (\alpha +\frac{2 (d-3) \phi ^2}{(d-2) x^2})^2}\\\nonumber
&\Bigl[&\frac{384 \pi  (d-3)^3 P \phi ^6}{(d-2)^3 x^6}-\frac{24 (d-3)^4 \phi ^6}{(d-2)^2 x^4}+(d-2) \\\nonumber
&(&\alpha ^2 (d-7)+\frac{12 (d-3)^3 \phi ^4}{(d-2)^2 x^4}+\frac{6 \alpha  (d-5) (d-3) \phi ^2}{(d-2) x^2})\Bigr],
\end{eqnarray}
 \begin{eqnarray}
\label{definitionentropylove}
S&=&\frac{2^{\frac{d}{2}-5} (d-2) x^2 \left(\frac{\sqrt{\frac{d-3}{d-2}} \phi }{x}\right)^d}{(d-6) (d-4) (d-3)^3 \phi ^6} \Bigr(4 (d-3)^2 \\\nonumber
&\times& (d^2-10 d+24) \phi ^4+\alpha ^2 (d-4) (d-2)^3 x^4\\\nonumber
&+&4 \alpha  (d-2)^2 \left(d^2-9 d+18\right) x^2 \phi ^2\Bigl)
 \end{eqnarray}
Thus, the equation of state for these black holes from Eq. (\ref{definitiontemlove}) in the grand canonical ensemble is given by
  \begin{eqnarray}
 \label{equationstatechlove}
  P&=&\frac{\sqrt{\frac{d-3}{d-2}} (d-2)^4 x^5 \left(\alpha +\frac{2 (d-3) \phi ^2}{(d-2) x^2}\right)^2}{16 \sqrt{2} (d-3)^3 \phi ^5}\\\nonumber
&&\Bigr(T-\frac{x}{12 \sqrt{2} \pi  \sqrt{\frac{d-3}{d-2}} \phi  \left(\alpha  (d-2) x^2+2 (d-3) \phi ^2\right)^2} \\\nonumber
&\times&\Bigl(6 \alpha  \left(d^3-10 d^2+31 d-30\right) x^2 \phi ^2\\\nonumber
&&+\alpha ^2 (d-7) (d-2)^2 x^4+12 (d-3)^3 \phi ^4\Bigr)\\\nonumber
&&+\frac{\sqrt{2} (d-3)^3 \sqrt{\frac{d-3}{d-2}} x \phi ^5}{\pi  \left(\alpha  (d-2) x^2+2 (d-3) \phi ^2\right)^2} \Bigl).
 \end{eqnarray}
The critical points should satisfy the following conditions
  \begin{eqnarray}
 \label{conditionchlove}
 \ \frac{\partial P}{\partial v}=\frac{{{\partial }^{2}}P}{\partial {{v}^{2}}}=0.
 \end{eqnarray}
 Considering Eqs. (\ref{equationstatechlove}) and (\ref{conditionchlove}), for $\phi =0.5$, $\alpha =1$ and $d=7$  we have only one critical point with positive pressure at $P_{c1}=0.022204$. \\
Also, we have the following two critical points for $\phi=0.5$, $\alpha =1$ and $d=8$ with a positive pressures at $P_{c1}=0.00909895$ and $P_{c2}=0.0476283$.\\

The Gibbs free energy in the grand canonical ensemble is given by
\begin{eqnarray}
\label{52b}
G =M-TS-Q\phi,
 \end{eqnarray}
We plot the diagram of the Gibbs free energy with respect to  temperature  using Eqs. (\ref{definitionmasslove}), (\ref{definitiontemlove}) and (\ref{definitionentropylove}), for $\alpha=1$ and $d=7$. The critical behaviors of these black holes for a given $\alpha$ shows  Van der Waals phenomena similar to the case of the third order of Lovelock-BI-AdS black holes in Fig.\ref{figure:Lovgrand}. In other words, we have the swallowtail behavior for $P<{{P}_{c}}$ and a second order phase transition occurs at $P={{P}_{c}}$. For $P>P_c$ there is no phase transition.\\
We have the critical points with the positive values of the Gibbs free energy for $\phi > 0.65$ (Fig. \ref{figure:Lovgrand}). By decreasing $\phi$, we observe the critical points for $\phi\leq 0.65$ with  negative values for the Gibbs free energy.\\
Also, we consider these calculations in $d=8$, we observe a reentrant phase transition for $\alpha=1$ and $\phi \leq 0.58$. This reentrant phase transition appears between two critical points $P_{c1}$ and $P_{c2}$, with the diagram being similar to the third order of Lovelock-BI-AdS black holes in Fig. \ref{figure:LovgrandRE}.  There is no phase transition for $\phi > 0.58$ and $\alpha=1$.\\
Consequently, we observe that the critical behaviors of the third order of Lovelock-charged-AdS black holes are similar to those of the third order of Lovelock-BI-AdS black holes and that their critical behaviors are independent of the parameter $\beta$ (Table. \ref{grandcanonic}).

Here, we consider the other options to calculate the heat capacity for  mixed ensemble with a fixed electric potential, pressure and ${\cal{A}}$ in the following form
 \begin{eqnarray}
  \nonumber & {{C}_{\phi ,P,{\cal{A}}}}({{r}_{+}},\phi ,P,{\cal{A}})=T{{(\frac{\partial S}{\partial T})}_{\phi ,P,{\cal{A}}}}=0,
 \end{eqnarray}\label{55b}
We find that there is no phase transition for the above heat capacity. Corresponding to this heat capacity, we can define a Grand potential $ {{\Omega }_{1}}({{r}_{+}},\phi ,P,{\cal{A}})=H-TS-Q\phi -\cal{A}\alpha$.\\
Moreover, for an ensemble  with fixed electric charges, $P$ and $\cal{A}$, one can obtain the heat capacity as follows:
  \begin{eqnarray}
 \label{59b}
 \nonumber & {{C}_{Q ,P,A }}({{r}_{+}},Q ,P,{\cal{A}})=T{{(\frac{\partial S}{\partial T})}_{Q ,P,{\cal{A}}}}=0,
 \end{eqnarray}
The thermodynamic potential   ${{\Omega }_{2}}({{r}_{+}},Q ,P,{\cal{A}})=H-TS-{\cal{A}} \alpha$  corresponds to the above heat capacity. Thus, there is no phase transition in these ensembles.

%%%%%%%%%%%%%%%%%%%%%%%%%%%%%%%%%%%%%%%%%%%%%%%%%%
 \section{Conclusion}

In this paper, we investigated the critical behaviors of the GB-BI-AdS black holes in the canonical (fixed $Q$) and grand canonical (fixed $\phi$) ensembles for $d=5,6$. We assumed the extended phase space with a cosmological constant and the coupling coefficient $\alpha$ and Born-Infeld parameter $\beta$ as the thermodynamic pressures of the system.\\
It was shown that the GB-BI-AdS black holes exhibit  interesting thermodynamic phenomena that depend on the coupling coefficient $\alpha$ in the canonical ensemble.
We also observed  "reentrant and triple point phase transitions" (RPT-TP) as well as  "multiple reentrant phase transitions" (multiple RPT) for different ranges of pressure depending   on the coefficient $\alpha$ in the canonical ensemble.
For $0 \leq \alpha <13$, the system was observed to behave  similar to the standard liquid/gas of the Van der Waals fluid. For $13 \leq \alpha <16$, $18 \leq \alpha <25$, and $32 \leq \alpha <40$, the black hole system admitted a reentrant large/small/large black holes phase transition. For $16 \leq \alpha <18$, a reentrant phase transition was found to occur for a specific range of pressure. Increasing  pressure led to a triple point for the special value of pressure. Also, two reentrant phase transition occurred for $25 \leq \alpha <32$. In other words, a reentrant large/small/large black holes phase transition occurred for a specific range of pressures and another reentrant phase transition happened by increasing  pressure. Finally, no phase transition occurred for $\alpha\geq 40$. \\
We also studied  the diagram of  critical pressures with respect to the coupling coefficient $\alpha$ of the GB-BI-AdS black holes. A minimum critical pressure was obtained at $\alpha=16$ at which  both the reentrant and triple point behaviors appeared.\\
GB-BI-AdS black holes  were considered in the grand canonical ensemble to find that they behave similar to the standard liquid/gas of the Van der Waals fluid for $d=5$ and to observe a reentrant phase transition  for $d=6$ and the specific value of $\phi$. Extending our calculations to the BI-AdS black holes for $d=5,6$, we observed the Van der Waals behavior for the given $d$.\\
The study of the critical behaviors of BI-AdS black holes in the third order of Lovelock gravity in the grand canonical ensemble revealed a  Van der Waals behavior for $d=7$ and a reentrant phase transition for $\phi \leq 0.6$ for $d=8$ at a specific range of pressure in the grand canonical ensemble. \\
Furthermore,  the limit of $\beta \to \infty$ was considered  for these black holes, i.e charged-AdS black holes in the third order of the Lovelock gravity in the grand canonical ensemble. Similar to the previous case, we observed a Van der Waals behavior for $d=7$ and a reentrant phase transition in $d=8$ for $\phi \leq 0.58$. Thus, the critical behaviors were shown to be  independent of the value of coefficient $\beta$. We also extended our calculations to different  mixed ensembles and the results showed no phase transitions in these ensembles.

\section{References}

%\begin{thebibliography}{000} %for 3 digits
%\begin{thebibliography}{00}  %for 2 digits

\end{document}